\documentclass[fleqn,usenatbib]{mnras}

\usepackage{newtxtext,newtxmath}

\usepackage[T1]{fontenc}
\usepackage{ae,aecompl}
\usepackage{float}

\usepackage{graphicx}	
\usepackage{amsmath}	
\usepackage{amssymb}	
\usepackage{bm}

\title[Testing the accuracy of ionospheric FR with LOFAR]{Testing the accuracy of the ionospheric Faraday rotation corrections through LOFAR 
observations of bright northern pulsars}
\author[]{N. K. Porayko$^{1}$\thanks{E-mail: nporayko@mpifr-bonn.mpg.de},
A. Noutsos$^{1}$,
C. Tiburzi$^{1,2}$,
J.P.W. Verbiest$^{2,1}$,
A. Horneffer$^{1}$,
\newauthor
J. K\"unsem\"oller$^{2}$,
S.  Os\l{}owski$^{3}$,
M. Kramer$^{1,4}$,
D.H.F.M. Schnitzeler$^{1}$,
J.M. Anderson$^{5}$,
\newauthor
M. Br\"uggen$^{6}$,
J.-M. Grie{\ss}meier$^{7,8}$,
M. Hoeft$^{9}$,
D.J. Schwarz$^{2}$,
M. Serylak$^{10,11,8}$,
\newauthor
O. Wucknitz$^{1}$
\\
\\
$^{1}$Max-Planck-Institut f\"ur Radioastronomie, Auf dem H\"ugel 69, D-53121 Bonn, Germany\\
$^{2}$Fakult\"at f\"ur Physik, Universit\"at Bielefeld, Postfach 100131, D-33501 Bielefeld, Germany\\
$^{3}$Swinburne University of Technology, PO Box 218, Hawthorn, VIC 3122, Australia\\Correction for the ionospheric Faraday rotation in LOFAR pulsar data
$^{4}$Jodrell Bank Centre for Astrophysics, School of Physics and Astronomy, The University of Manchester, Manchester M13 9PL, UK\\
$^{5}$Deutsches GeoForschungsZentrum GFZ, Telegrafenberg, 14473 Potsdam, Germany\\
$^{6}$Universit\"at Hamburg, Hamburger Sternwarte, Gojenbergsweg 112, D-21029, Hamburg, Germany\\
$^{7}$LPC2E - Universit\'{e} d'Orl\'{e}ans /  CNRS, 45071 Orl\'{e}ans cedex 2, France\\
$^{8}$Station de Radioastronomie de Nan\c{c}ay, Observatoire de Paris, PSL Research University, CNRS, Univ. Orl\'{e}ans, OSUC, 18330 Nan\c{c}ay, France\\
$^{9}$Th\"uringer Landessternwarte, Sternwarte 5, 07778 Tautenburg\\
$^{10}$SKA SA, 3rd Floor, The Park, Park Road, Pinelands, 7405, South Africa\\
$^{11}$Department of Physics and Astronomy, University of the Western Cape, Cape Town 7535, South Africa\\
}

\date{Accepted XXX. Received YYY; in original form ZZZ}

\pubyear{2018}

\begin{document}
\label{firstpage}
\pagerange{\pageref{firstpage}--\pageref{lastpage}}
\maketitle

\begin{abstract}
Faraday rotation of polarized emission from pulsars measured at radio frequencies provides a powerful tool to investigate the interstellar and interplanetary magnetic fields. However, besides being sensitive to the astrophysical media, pulsar observations in radio are affected by the highly time-variable ionosphere. In this article, the amount of ionospheric Faraday rotation has been computed by assuming a thin layer model. For this aim, ionospheric maps of the free electron density (based on Global Positioning System data) and semi-empirical geomagnetic models are needed. Through the data of five highly polarized pulsars observed with the individual German LOw-Frequency ARray stations, we investigate the performances of the ionospheric modelling. In addition, we estimate the parameters of the systematics and the correlated noise generated by the residual unmodelled ionospheric effects, and show the comparison of the different free-electron density maps. For the best ionospheric maps, we have found that the rotation measure corrections on one-year timescales after subtraction of diurnal periodicity are accurate to $\sim$ 0.06--0.07 rad m$^{-2}$.
\end{abstract}

\begin{keywords}
pulsars: general--stars:neutron--polarization--atmospheric effects
\end{keywords}



\section{Introduction}\label{sec:intro}
Since their discovery \citep{1968Natur.217..709H}, pulsars have been a powerful tool to probe the magnetoionic plasma. Due to frequency-dependent dispersion delay and scattering of their signals, pulsars can be used to study, e.g., turbulence in the ionised interstellar medium (ISM) on many orders of magnitude \citep[e.g.][]{1977ARA&A..15..479R, 1995ApJ...443..209A, 2007MNRAS.378..493Y}, the distribution of free electrons in the Milky Way and the Local Bubble \citep[e.g.][]{2002astro.ph..7156C, 1998ApJ...500..262B}, and the electron content of the Solar wind \citep[e.g.][]{2007ApJ...671..907Y, 2016ApJ...831..208H}. Magnetised plasma also induces Faraday rotation in linearly polarised radiation, that is, a rotation of the polarization angle $\psi$ depending on the radiation wavelength $\lambda$,
\begin{equation}
\psi=\psi_0+\text{RM}\text{ } \lambda^2,
\end{equation}

\noindent where RM is the rotation measure defined as:
\begin{equation}
\text{RM} = 0.81 \int^{\text{observer}}_{\text{source}} n_{\text{e}}\bm{B} \cdot d\bm{r}~[\text{rad}\ \text{m}^{-2}],
\label{RMlambda}
\end{equation}
with $n_{\text{e}}$ being the electron density in the ionised ISM $[\text{cm}^{-3}]$, $\bm{B}$ the magnetic field $[\mu\text{G}]$ and $d\bm{r}$ [pc] the infinitesimal interval of the distance along the line of sight (LoS). From the above expressions one can see that more accurate RM estimations can be achieved with broad-band instruments operating at longer wavelengths.

Due to the high percentage of linear polarisation, and low levels of magnetospheric Faraday rotation \citep[e.g.][]{2011MNRAS.417.1183W}, pulsars are useful objects to measure RM induced by the ionised ISM, and hence the Galactic magnetic fields \citep[e.g.][]{2018ApJS..234...11H}. 

Because the propagation effects  are strongly depend on $\lambda$, low frequencies are favoured for studies of these effects in pulsars\footnote{For strongly Faraday rotated sources, such as pulsars in the dense regions(e.g. magnetar in the Galactic
center) and distant active
galactic nuclei, RMs can be as well effectively probed with instruments, operating at 1-2 GHz and higher frequencies.}. Moreover, the steep spectra of pulsars \citep[e.g.][]{2013MNRAS.431.1352B} and the reduction of the linear polarization fractions at high frequencies in pulsar emission \citep[e.g.][]{2008MNRAS.388..261J}, make the low-frequency band even more preferable for Faraday rotation studies.

Recently, low-frequency pulsar astronomy was revived thanks to a number of cutting-edge facilities such as the Long Wavelength Array (LWA) \citep{2009IEEEP..97.1421E}, the Murchison Widefieled Array (MWA) \citep{2013PASA...30....7T}, the Giant Ukrainian Radio Telescope (GURT) \citep{2016JAI.....541010Z} and the LOw Frequency ARray (LOFAR) \citep{2013A&A...556A...2V, 2011A&A...530A..80S}.

Nevertheless, polarisation studies at low frequencies are challenging. Besides the effects of the magnetised ionised ISM, linearly polarised radiation can be noticeably rotated by the highly variable terrestial ionosphere. Moreover, it can significantly depolarise observations when averaging over several hours. For a review on the propagation of radio waves through the ionosphere, see e.g. \citet{2013tra..book.....W} or \citet{2001isra.book.....T}. 

In order to mitigate the ionospheric contribution to FR, numerous techniques have been developed. One very promising approach is based on providing quasi-simultaneous observations of a known background source (e.g. the diffuse polarised background), located within the ionospheric correlation spatial scale with respect to the source of interest, to recover the ionospheric Faraday rotation \citep{2016ApJ...830...38L}. 

Alternatively, the ionospheric Faraday rotation can be estimated by combining models of the ionospheric electron density and of the geomagnetic field. In the majority of the studies that aimed to measure the interstellar Faraday rotation in astronomical sources \citep[e.g.][]{2004ApJS..150..317W, 2006ApJ...642..868H, 2011MNRAS.414.2087Y}, the ionospheric electron density was computed through the semi-empirical International Reference Ionosphere (IRI) model \citep{2014JSWSC...4A..07B}, which provides monthly-averaged ionospheric electron density profiles up to 2000 km, as a function of time and location. However, due to the sparsely distributed ground and space observatories that contribute to the IRI model, and the large averaging time, the modelled values of electron densities can significantly deviate from the real ones \citep{2007AdSpR..39..841M}.
Higher accuracies can be reached by a technique described in \citet{2001A&A...366.1071E}, where the ionospheric electron densities are obtained through raw dual-frequency GPS data, recorded with a set of local GPS receivers. When applied to PSR~J1932+1059, the variance of the differences between the observed RM as obtained at the VLA, and the predicted ionospheric RM as computed with the AIPS APGPS routine\footnote{A similar approach is implemented in the ALBUS software \url{https://github.com/twillis449/ALBUS_ionosphere}}, was found to be 0.2 rad m$^{-2}$. 

A handier and less computationally expensive alternative to this approach consists in using global ionospheric maps of electron column densities in the ionosphere, which are based on the available data from all the GPS stations spread around Earth. This technique was implemented and tested on a set of pulsars by \citet{2013A&A...552A..58S}, showing a qualitatively good agreement between the expected and the observed values of FR. However, \citet{2013A&A...552A..58S} have restricted their analysis to probing only two global ionospheric maps (ROBR and CODG), and the research was carried out on a set of observations with timespans of only several hours. The standard deviations of the residuals between the RMs, observed and modelled with CODG and ROBR, varied for different datasets in the ranges 0.12--0.20 rad m$^{-2}$ and 0.07--0.20 rad m$^{-2}$, respectively.

In this article, we aim to compare the performance, and estimate the accuracy of different publicly available global ionospheric maps, when applied to correct for ionospheric Faraday rotation in several months of pulsar data. For these goals, we used pulsar observations obtained with the international LOFAR stations in Germany. In Section~\ref{sec:obs} we describe the instrumental and observational setup and our data reduction, including a first, application of a simple ionospheric modelling. 
In Section~\ref{sec:system_in_shortdata} we attempt to model the ionospheric Faraday rotation in our dataset and we analyse the systematics left in the RM residuals after ionospheric mitigation. In Sections \ref{sec:below1yr}, \ref{subsec:linear} we focus on how to correct for the systematics, and show the results obtained after the implementation of our additional corrections and the comparison of different global ionospheric maps. In Section \ref{sec:sum} we then summarise our findings. 


\section{Observations and Data Reduction}
\label{sec:obs}
LOFAR, the LOw-Frequency ARray, is an international interferometric telescope operating at very low frequencies, from 10 up to 240 MHz \citep{2013A&A...556A...2V}. LOFAR stations are distributed throughout Europe, with a dense core, the Superterp, located in the Netherlands. Six of the stations are located in Germany: DE601 in Effelsberg, DE602 in Unterweilenbach, DE603 in Tautenburg, DE604 in Bornim, DE605 in J\"ulich and DE609 in Norderstedt. 

Each LOFAR station outside the Netherlands consists of 96 pairs of dipoles of low band antennas (LBAs) and 96 tiles (i.e., groups formed of 16 pairs of dipoles) of high-band antennas (HBAs). Typically three days a week the German stations are used as stand-alone telescopes by the GLOW (German Long Wavelength) consortium\footnote{GLOW is an association of German universities and research institutes, which promotes the use of the meter wavelength spectral window for astrophysical purposes. GLOW members operate the work of the German LOFAR stations and GLOW is further involved at the planned Square Kilometer Array (SKA) project. More details can be found in \url{https://www.glowconsortium.de/index.php/en/}}, to perform an observing campaign of pulsars at low frequencies ($\sim$100--200 MHz) using the HBAs. Commonly, each German station in stand-alone mode observes a unique set of pulsars with $\sim$2-hours integration time per pulsar. 
	The specifics for the dataset used in the presented analysis are summarised in Table \ref{table1}. As it will be shown in the next sections, we focus our analysis on the characterization of potential short- and long-term trends in the residuals between the ionospheric models and the data. All the selected pulsars have high signal-to-noise ratio (S/N), which varies from $\sim$800 up to $\sim$2000, and a significant fraction of linear polarization (at least 10\%), which allows us to measure the RM with high accuracy and precision. Besides this, for the purposes of the short-term analysis, we chose pulsars with a significant fraction of long, continuous observations (from a few hours to entire days). This allows us to properly identify also high-frequency systematics. For the purposes of the long-term analysis, this last requirement is not strictly necessary, and we thus selected pulsars with a long observing baselines.

\begin{table}
\caption{Details of the observations used for the white noise plateau investigation (see Sec. \ref{sec:below1yr}) and for the long-term systematics (see Sec. \ref{subsec:linear})}
\label{table1}
Short-term
\begin{tabular}{lcccr}
\hline
\hline 
Jname & Site & $T_{\text{obs}}$  \\
\hline
J0332+5434 & DE609 &  2015-12-19 -- 2016-06-13 \\
J0814+7429 & DE605 &  2016-01-08 -- 2017-04-30 \\
J1136+1551 & DE601 &  2016-01-09 -- 2016-10-09 \\
\hline
\end{tabular}
Long-term
\begin{tabular}{lcccr}
\hline
\hline 
Jname & Site & $T_{\text{obs}}$  \\
\hline
J0332+5434 & DE605 &  2014-03-09 -- 2017-02-11 \\
J0826+2637 & DE603 &  2015-02-22 -- 2017-02-03 \\
J1136+1551 & DE601 &  2013-09-06 -- 2016-12-31 \\
J1921+2153 & DE605 &  2014-03-08 -- 2017-02-11 \\
\hline
\end{tabular}
\end{table}

After digitizing and beamforming, the data have 5.12$\mu$s time resolution and are split into frequency channels of 195 kHz bandwidth. Due to data rate limitations, only 366 (488 for DE601) channels are recorded on machines at the Max-Planck-Institut f\"ur Radioastronomie in Bonn and at the J\"ulich Supercomputing Centre using the \textsc{LuMP} (LOFAR und MPIfR Pulsare) Software\footnote{\url{https://github.com/AHorneffer/lump-lofar-und-mpifr-pulsare}}. The datasets are then, coherently de-dispersed, folded modulo the pulse period and reduced to more manageable 10-second sub-integrations with the \textsc{DSPSR} software\footnote{\url{http://dspsr.sourceforge.net/}} \citep{2011PASA...28....1V}, and stored as \textsc{PSRFITS} archives \citep{2004PASA...21..302H}. We then excise the radio-frequency interference with the \textsc{COASTGUARD}'s clean.py surgical algorithm \citep{2016MNRAS.458..868L}.

In contrast to steerable radio telescopes, the LOFAR antennas are fixed on the ground, which causes a distortion of the polarisation signal, as well as decrease of the intrinsic signal intensity, towards low elevations, due to the projection effects \citep{2015A&A...576A..62N}. 
For instance, such an instrumental response is responsible for the so-called instrumental peak at $0$ rad m$^{-2}$ in the RM spectrum while performing the RM synthesis analysis \citep{1966MNRAS.133...67B, 2005A&A...441.1217B}.
We mitigate these instrumental effects by applying a Jones calibration matrix based on the Hamaker measurement equations \citep[see][]{1996A&AS..117..137H, 2011A&A...527A.106S}. However, \citet{2015A&A...576A..62N} showed that across several hours of observations taken with the Superterp, the intrinsic signal intensity of the LOFAR antennas significantly degrades at low elevations ($\lesssim$ $30\degr$) even after  the calibration procedure has been applied.

Due to the fact that radio observations in the LOFAR frequency band are quite sensitive to the highly variable ionospheric layer \citep{2016ApJS..223....2V}, we split pulsar archives into 15-minute subintegrations with the \textsc{PSRCHIVE} software package\footnote{\url{http://psrchive.sourceforge.net/}} \citep{2012AR&T....9..237V}, which corresponds to the minimum time-sampling of ionospheric maps that we have tested (see Section \ref{sec:model_ion}). This reduces the unresolved contribution of ionospheric RM, while still providing a reasonable S/N. 

After this, we estimate the RM for each of the 15-minute subintegrations, building an RM time series for each of the analysed datasets. For this, we use an optimised version of the classical RM synthesis technique, described in Appendix~\ref{app:rm}. 

\begin{figure}
\includegraphics[width=1.\columnwidth]{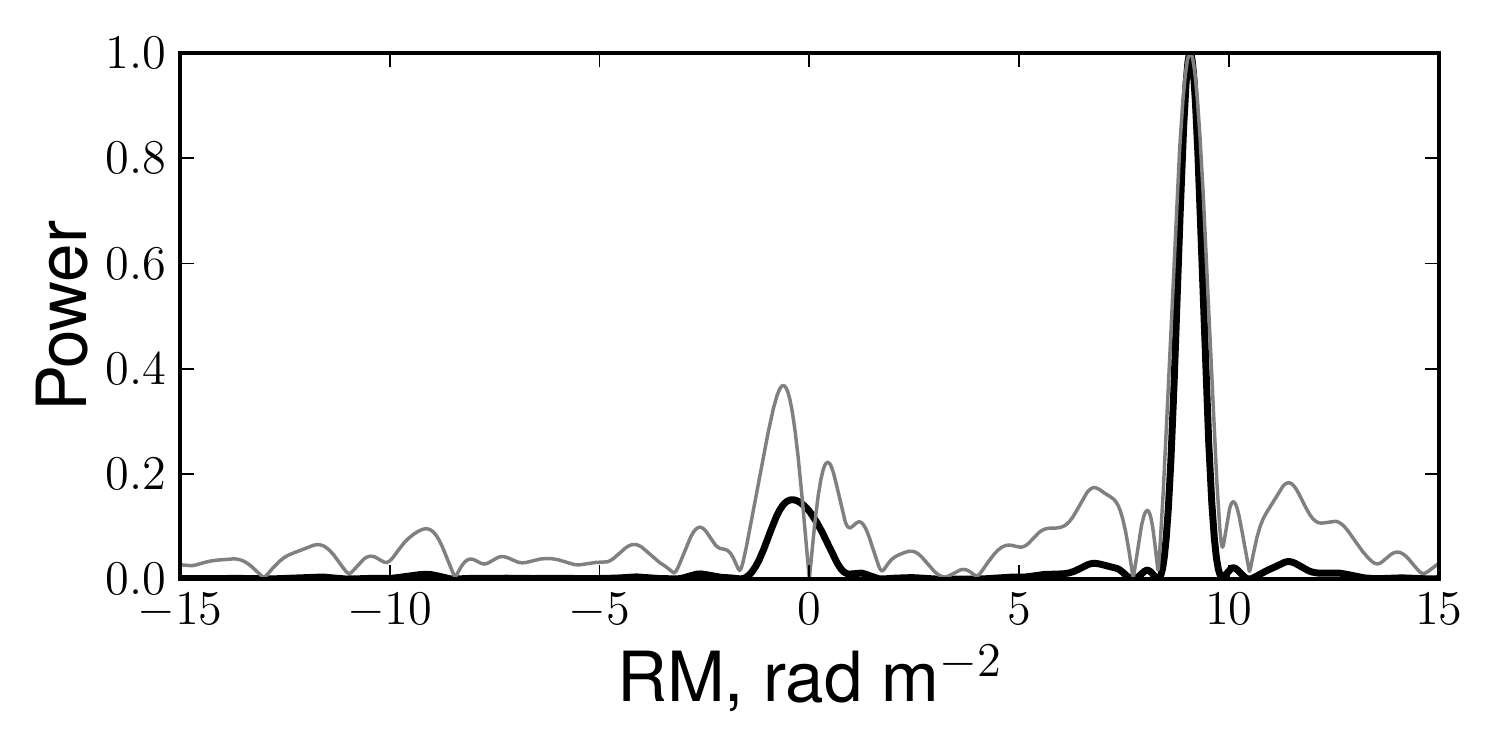}
\includegraphics[width=1.\columnwidth]{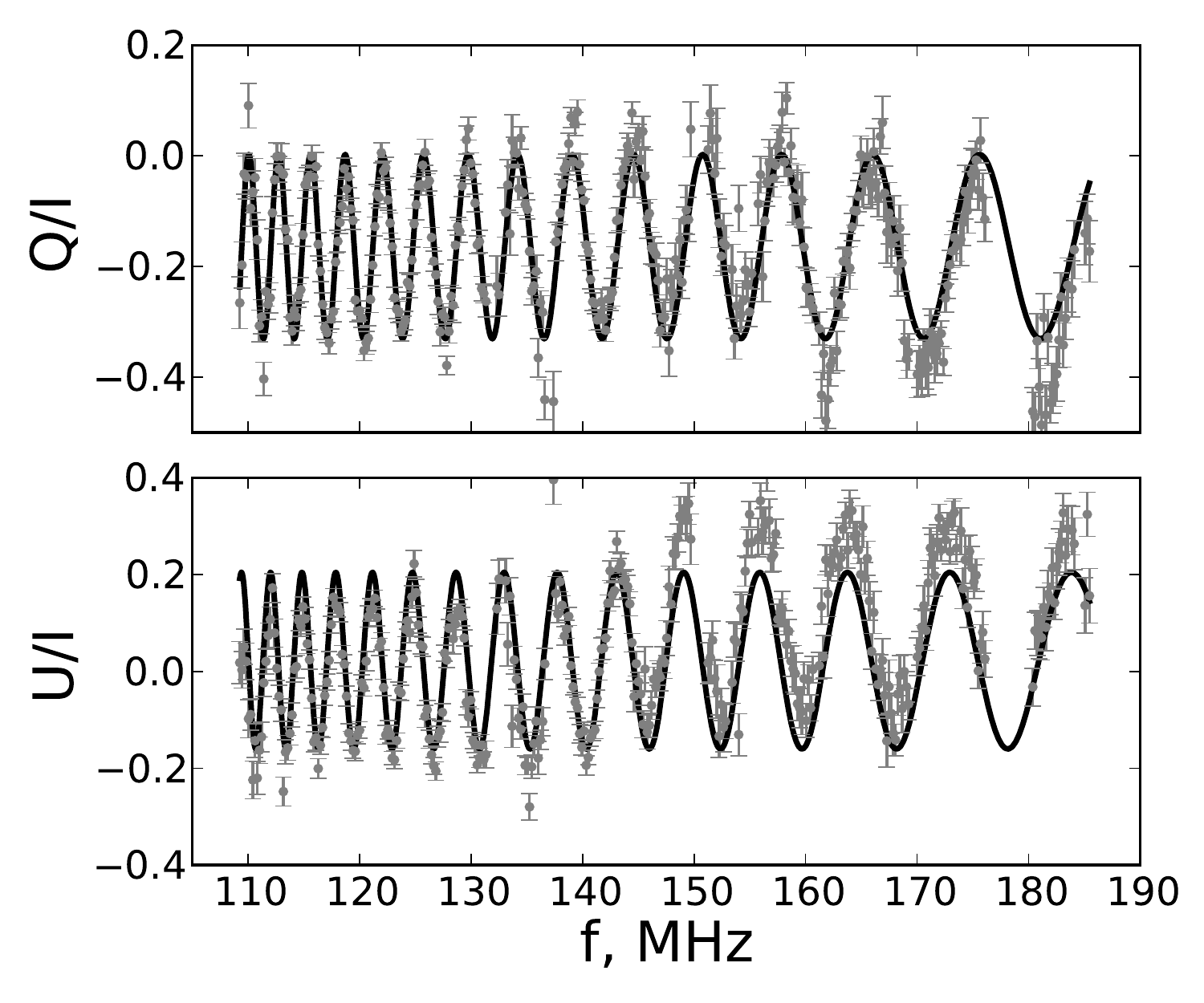}
\caption{\textit{Top panel} -- A comparison between the RM spectrum obtained with the classical RM synthesis (grey line) and the logarithm of the RM posterior probability (black line) given by Eq.~(\ref{Ppost}) for a 15-minute observation of PSR~J1136+1551. All the curves are normalised to the maximum values. The maximum peak corresponds to the observed pulsar RM=$9.076$ rad m$^{-2}$. \textit{Bottom panel} -- Harmonic variations of the Stokes parameters $Q$ and $U$ across the observed bandwidth (grey points). The black lines show the expected harmonic trend, given the pulsar's RM.}
    \label{fig:RMsynt}
\end{figure}

\begin{figure}
\includegraphics[width=1.1\columnwidth]{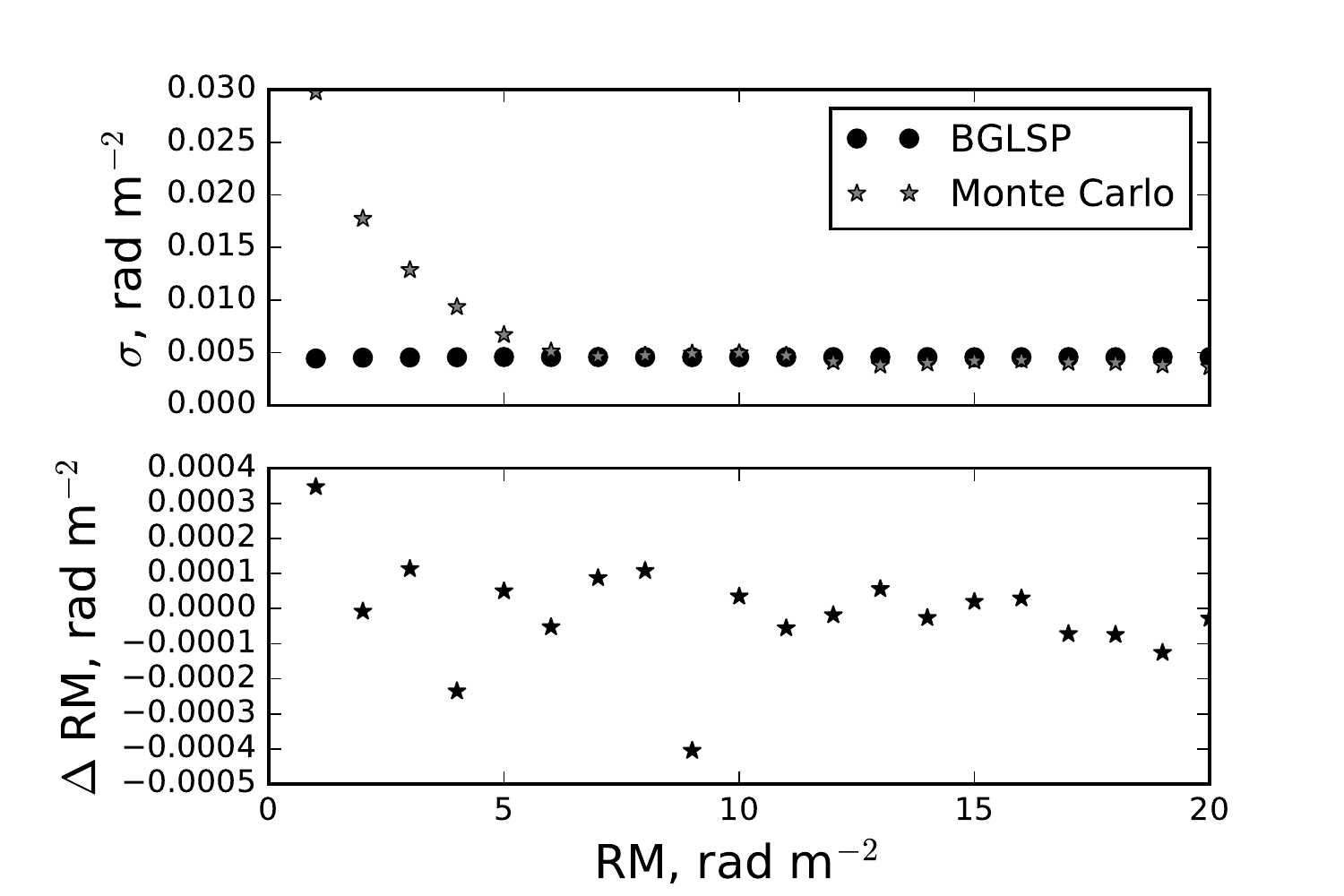}
    \caption{\textit{Top panel} -- The uncertainties on the RM values detected with BGLSP (black circles), overplotted with the variance of the distribution of the detected RMs obtained from Monte Carlo simulations (grey stars). The plot demonstrates that BGLSP uncertainties are underestimated for |RM|<6 rad m$^{-2}$. \textit{Lower panel} -- The difference between the injected RMs and the mean values of the Monte Carlo distributions. No systematic deviations between BGLSP and Monte Carlo can be seen. For both panels the reduced $\chi^2$ of the $u$ and $q$ fit was 10.}
    \label{fig:chi10}
\end{figure}

In Fig. \ref{fig:RMsynt} we demonstrate the example of the Bayesian Generalised Lomb-Scargle Periodogram (BGLSP) application to one of the 15-min observation of PSR J1136+1551. We clearly see  systematic deviation from the modelled $Q$ and $U$, which is reflected in the spectrum as a low-frequency excess of power around 0 rad/m$^2$. The origin of these systematics is not known for certain, but it is highly likely that it is associated with instrumental properties, e.g. non-linearity in the instrumental setup. Because the spurious peak affects a small range of values around 0 rad/m$^2$, we expect sources with significant larger RMs to be uneffected, suggesting little or no influence on our results. However, we point out that the results can be biased when dealing with astronomical sources with low RM values. In order to prove these considerations, we have performed two tests. Firstly, we tested the basic assumption that any discrepancies between the models and the data are induced by an effect that is strongly frequency dependent. Therefore, we have split data into two sub-bands and  measured RM values separately for the bottom and upper half of the bandwidth. The results show that both RM values are in excellent agreement within the uncertainties. This suggests that the effect is not strongly depending on frequency.
Still, we also tested whether a systematic effect could conspire to mimic a wrong RM value. As a worst case scenario, 
we have investigated the impact of systematics, in case they had a quadratic dependency on frequency, which would mimic the $\lambda^2$ dependency introduced by the physical effect of Faraday rotation. The simulated Stokes Q and U were evenly sampled in frequency with a realistic 20\% of data loss due to radio-frequency interference. We run a Monte Carlo simulation with $10^3$ realizations of this set-up, for increasing values of RMs from 0 to 20 rad/m$^2$. A range of the systematic amplitudes were tested with reduced $\chi^2$ of up to 10, as the reduced $\chi^2$ detected in the data did not exceed this value. We found that, starting from an RM value of $\sim\!6$ rad/m$^2$, the mean and variance of the distribution of the recovered RMs are in a good agreement with the results from BGLSP (see Fig. \ref{fig:chi10}). This behaviour is expected, since as soon as the source RM is larger than the width of the systematic feature, the two signals can be separated reliably.

With the reliability of our RM measurements established by these tests, we proceed to do a first attempt to mitigate the Faraday rotation ionospheric contribution.

\subsection{On modelling the ionospheric RM variations: thin layer ionospheric model}\label{sec:model_ion}

If not taken into account, the ionosphere introduces noise in the measured RM values. This makes it impossible, for instance, to investigate RM variations caused by the turbulent ionised ISM, which are expected to be $\sim$3--4 orders of magnitude lower than the root-mean-square (rms) of the ionospheric RM fluctuations (see Eq.~(\ref{RM-ism_pred})). We now briefly recap the ionospheric RM behavior and the ways to model it.

The ionospheric layer, partially consisting of free electrons and positively charged ionised molecules and atoms, extends from 50 km to beyond 2000 km above the Earth's surface \citep{1969itip.book.....R}. The ionospheric contribution to RM can be estimated to be of the order of 1--4 rad m$^{-2}$, however, the essential complexity in treating the ionospheric RM comes from its strong variability, which typically changes during the day up to 80$\%$. The ionization fraction of the ionospheric shell, mostly caused by photoionization processes involving the Sun's extreme ultra-violet and X-ray emission, varies significantly over timescales of minutes (due to Solar flares) up to years (11-year Solar cycle). Besides this, the ionosphere shows diurnal (caused by the relative motion of the Sun on the celestial sphere) and 27-day periodicities (due to the Solar rotation). As the Earth's atmosphere is not homogeneous and different molecules are dominating at different heights, the ionospheric shell, does not have a homogeneous electron density distribution, and achieves its maximum during the day time in the so-called F sublayer, which implies $\sim$ 50-60 \% of all the electrons in the ionosphere \citep{2017SpWea..15..418B}.
\begin{figure}
\includegraphics[width=1.1\columnwidth]{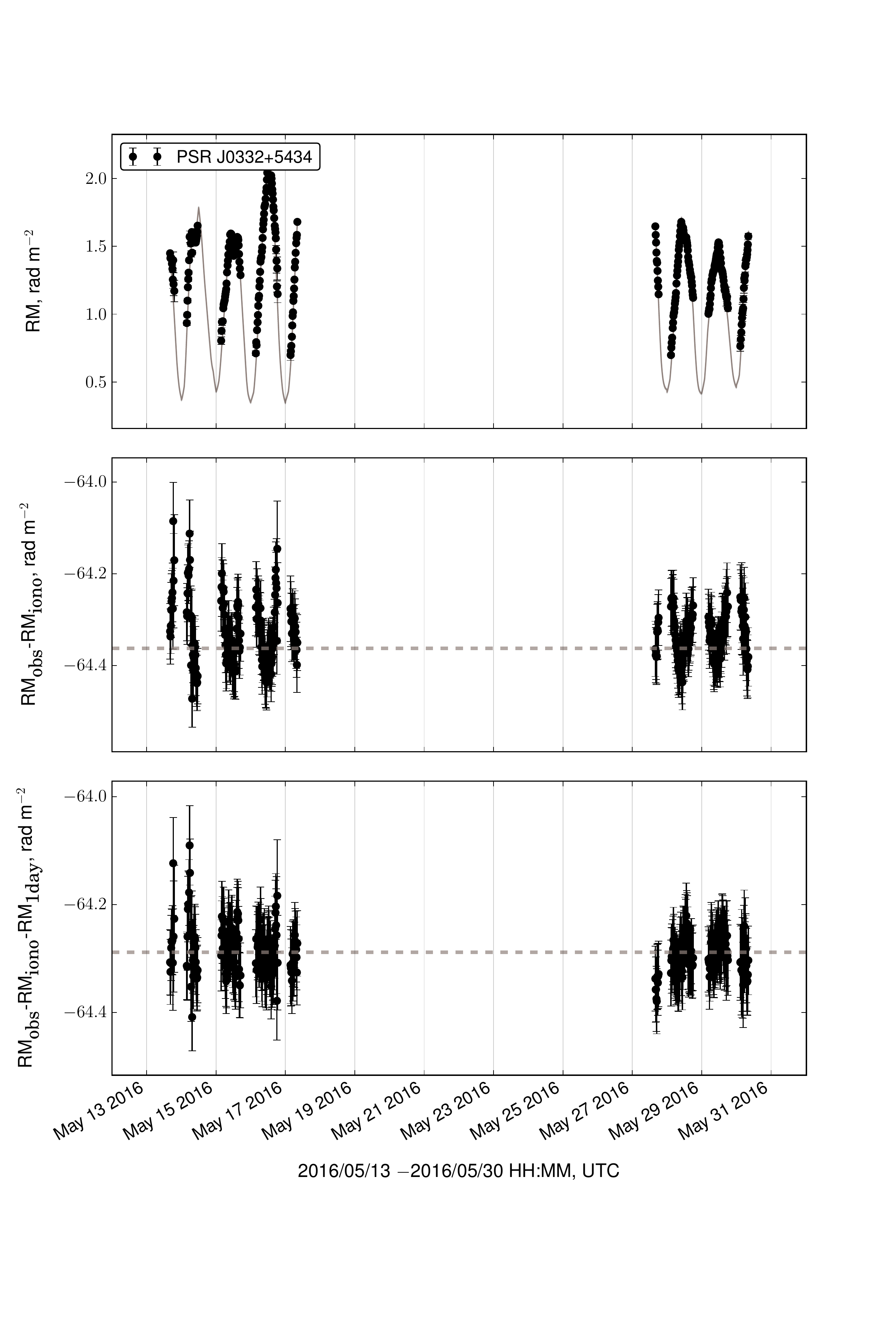}
    \caption{Example of application of JPLG ionospheric maps and POMME10 geomagnetic model to real data of PSR~J0332+5434 observed by DE609. \textit{Upper panel} -- modelled ionospheric RM computed with \textsc{RMextract} using JPLG ionospheric maps (in grey), applied to measured RMs shifted by a constant value $\text{RM}_{\text{IISM}}$ (black dots). The uncertainties on the modelled RM are smaller than the symbol used. \textit{Middle panel} -- residuals between observed and modelled RM (black dots) before subtraction of 1-day sinusoid. \textit{Lower panel} -- residuals between observed and modelled RM (black dots) after subtraction of 1-day sinusoid. The grey dashed line shows the constant value $\text{RM}_{\text{IISM}}$. The uncertainties on the measured RMs are modified by the values determined through the analysis described in Sec. \ref{sec:below1yr}. Only the observations above $\sim$ $30\degr$ in elevation were used.}
    \label{fig:variat}
\end{figure}
Because of this, the ionosphere can be reasonably well modelled by a thin shell located at the \textit{effective ionospheric height}, which is usually estimated to be between 300 and 600 km above the Earth's surface. 

As the projected thickness of the non-uniform ionospheric layer increases out from the zenith, it is common practice to discard data at low elevations. For this work we have used a $30\degr$ elevation cut-off\footnote{This number is partially motivated by the limitations of the polarization calibration method used in this work}. In the case of the ionosphere, and with the mentioned assumptions, Eq.~(\ref{RMlambda}) is reduced to \citep{2013A&A...552A..58S}:
\begin{equation}
\text{RM}_{\text{iono}}=2.6\times 10^{-17}\text{STEC} \times \textit{B}_{\text{iono}}\text{ rad m}^{-2},
\end{equation}
where STEC (Slant TEC, where TEC stands for `Total Electron Content') is equal to the column density of electrons [m$^{-2}$] at the cross-section between the LoS and the ionospheric shell and $B_{\text{iono}}$ is the projection of the magnetic field [G] in the F-layer on the LoS. The thin layer approximation has already been implemented in several codes aimed at the estimation of the ionospheric RM \citep[e.g. in][]{2013A&A...552A..58S}. In particular, for the work presented here we use the publicly available RMextract software\footnote{\url{https://github.com/lofar-astron/RMextract}}, that estimates the ionospheric RM along a certain LoS and at a certain point in time making use of a geomagnetic field model and a global ionospheric map. An example of ionospheric RM calibration with RMextract, applied to the RM sequence of PSR J0332+5434, is demonstrated on Fig. \ref{fig:variat} (upper panel). From here on in this paper for demonstration purposes we have used JPLG maps, which have showed the second best result in our analysis and are commonly available for the majority of our observing epochs.

\begin{figure}
\includegraphics[width=1.
\columnwidth]{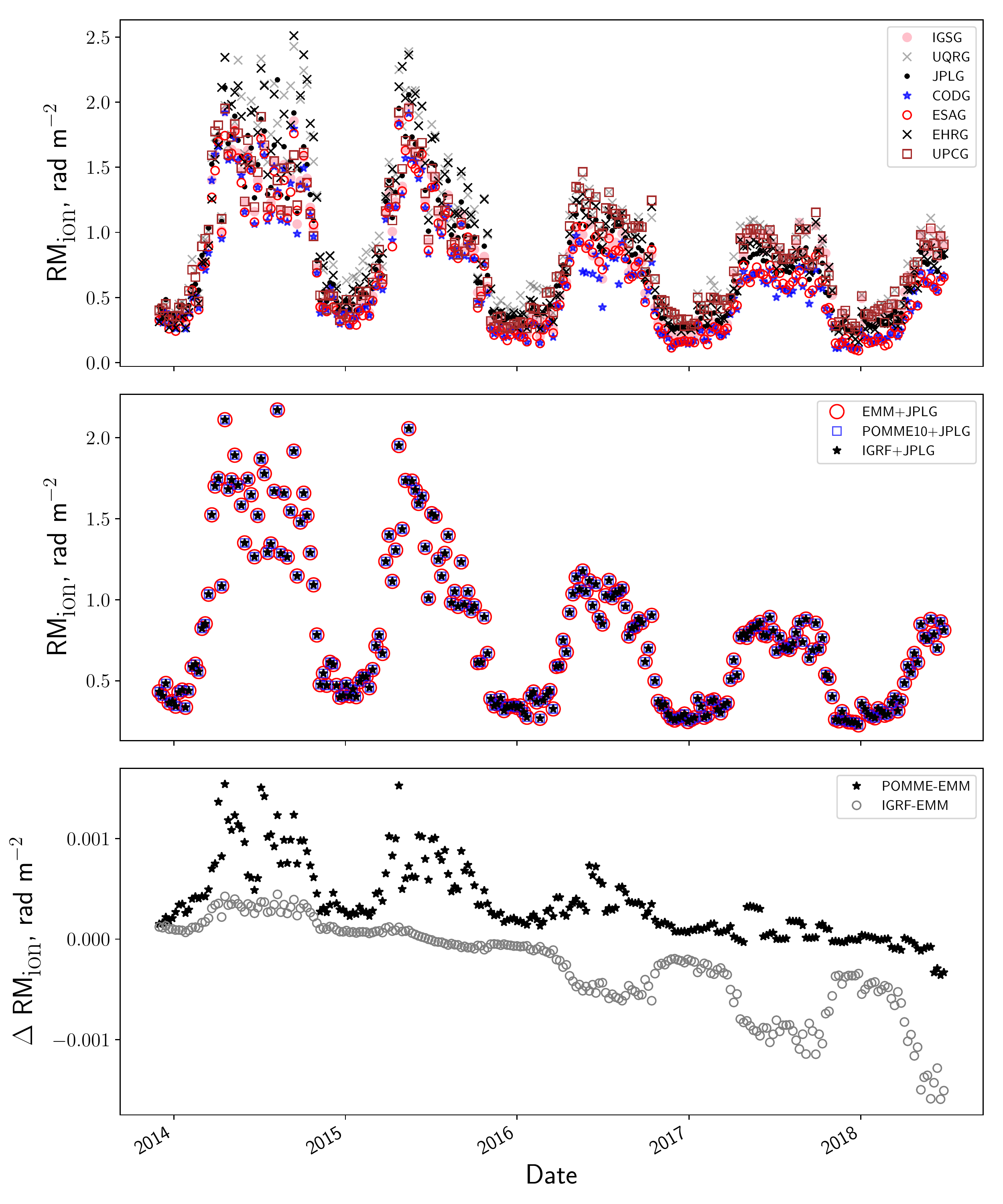}
\caption{(Colours online)\textit{Upper panel:} Comparison between ionospheric RMs in the direction of PSR~J0332+5434 observed at constant 30\degr~ 
 elevation, as modelled by different ionospheric maps (+POMME10 geomagnetic model). \textit{Middle panel:} Comparison between ionospheric RMs in the direction of PSR~J0332+5434 observed at constant 30\degr~ 
 elevation, as modelled by POMME10, EMM and IGRF12 (+JPLG ionospheric map). \textit{Lower panel:} Difference between ionospheric RMs in the direction of PSR~J0332+5434 observed at constant 30\degr~  elevation, as modelled by POMME10, EMM and IGRF12 (+JPLG ionospheric map). The empty circles show the difference between IGRF12 and EMM. The black stars show the difference between EMM and POMME10, which is on average less than 0.001 rad m$^{-2}$ for observations above 30\degr~ in elevation.}
    \label{fig:magn_pict1}
\end{figure}

\begin{figure}
\includegraphics[width=1.1\columnwidth]{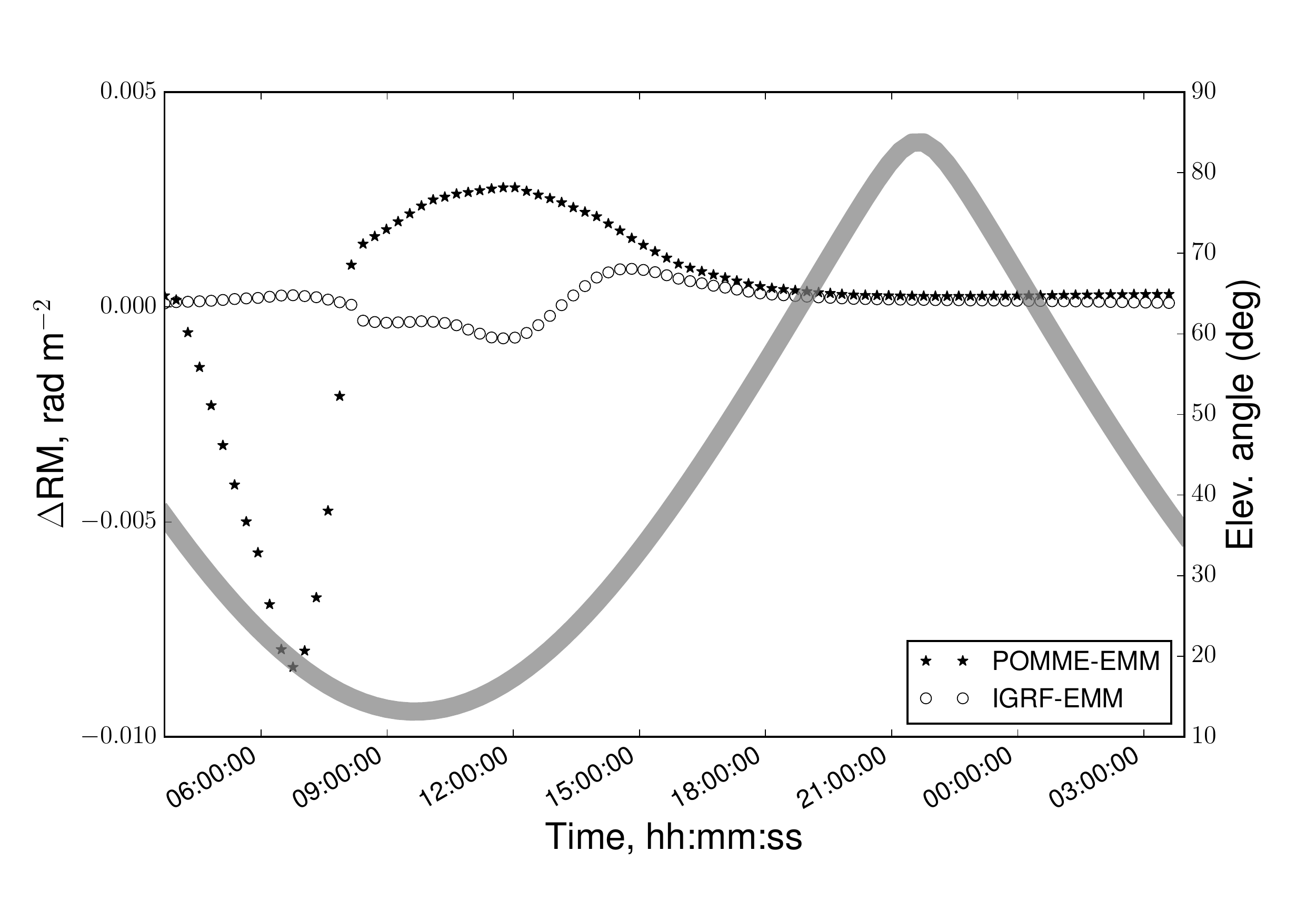}
\caption{Difference between ionospheric RMs in the direction of PSR~J0332+5434 observed at multiple elevations along the day, as modelled by POMME10, EMM and IGRF12. The empty circles show the difference between IGRF12 and EMM. The black stars show the difference between EMM and POMME10, which is on average less than 0.001 rad m$^{-2}$ for observations above 30\degr~ in elevation. The thick gray line shows the change in PSR~J0332+5434 elevation angle.}
    \label{fig:magn_pict2}
\end{figure}

The geomagnetic field models are conventionally represented as spherical harmonical expansions of a scalar magnetic potential. Several geomagnetic models are publicly available, among which are the Enhanced Magnetic Model (EMM)\footnote{\url{https://www.ngdc.noaa.gov/geomag/EMM/}}, the International Geomagnetic Reference Field (IGRF) \citep{2015EP&S...67...79T}, the World Magnetic Model\footnote{\url{https://www.ngdc.noaa.gov/geomag/WMM/DoDWMM.shtml}}, and POMME10\footnote{\url{http://geomag.org/models/pomme10.html}}. The lower panel of Fig. \ref{fig:magn_pict1} shows a comparison of the ionospheric magnetic field given by EMM, POMME10 and IGRF12, for the years 2013 through 2018 for lines of sight from Germany in the direction of 30\degr in elevation (minimum elevation used in our work). The plot demonstrates clear systematic behaviour, although, on average between 2013 and 2018, there is less than 0.1\% difference between different geomagnetic models. The discrepancy seems to be increasing with time. Thus, for the future datasets taken around 2020 geomagnetic models with non-evolving with time geomagnetic parameters will reach few per cent level difference between them and should be used with care. Fig. \ref{fig:magn_pict2} demonstrates that for low elevation observations this difference can hit 1\% from the absolute value.

We have conducted a full analysis by making use of all three geomagnetic models. In order to be concise, we present only the results of  POMME10 \citep{2006GGG.....7.7008M} here (see Table \ref{tab:example_table1}). In the case one of the other two geomagnetic models the results on parameter estimation and the presence of various systematics in the data remain unchanged, and are presented in Supplementary material online (according to Table \ref{tab:example_table1}).

The global ionospheric maps (publicly available\footnote{\url{ftp://cddis.gsfc.nasa.gov/gnss/products/}}$^,$\footnote{{\url{ftp://igs.ensg.ign.fr/}}}) in IONEX\footnote{\url{https://igscb.jpl.nasa.gov/igscb/data/format/ionex1.pdf}} format, provide estimates of the vertical TEC. Several global ionospheric maps are available: CODG (from the University of Bern), ESAG and EHRG (European Space Agency), JPLG (Jet Propulsion Laboratory), UPCG and UQRG (Technical University of Catalonia, see \citet{2005JASTP..67.1598O}), IGSG (International GNSS Service). Although the maps can be based on the same GPS data, the published TEC values can vary from group to group because of different interpolation schemes and different spatial and temporal resolution. In practice, the maps we have used, all have a spatial resolution of $2.5\degr \times 5\degr$(latitude $\times$ longitude). CODG and EHRG have a time resolution of 1 hour, UQRG of 0.25 hours and the remaining maps (ESAG, IGSG, JPLG and UPCG) have a time resolution of 2 hours.

\section{Systematics in the RM residuals}
\label{sec:system_in_shortdata}
\begin{figure}
\includegraphics[width=1.1\columnwidth]{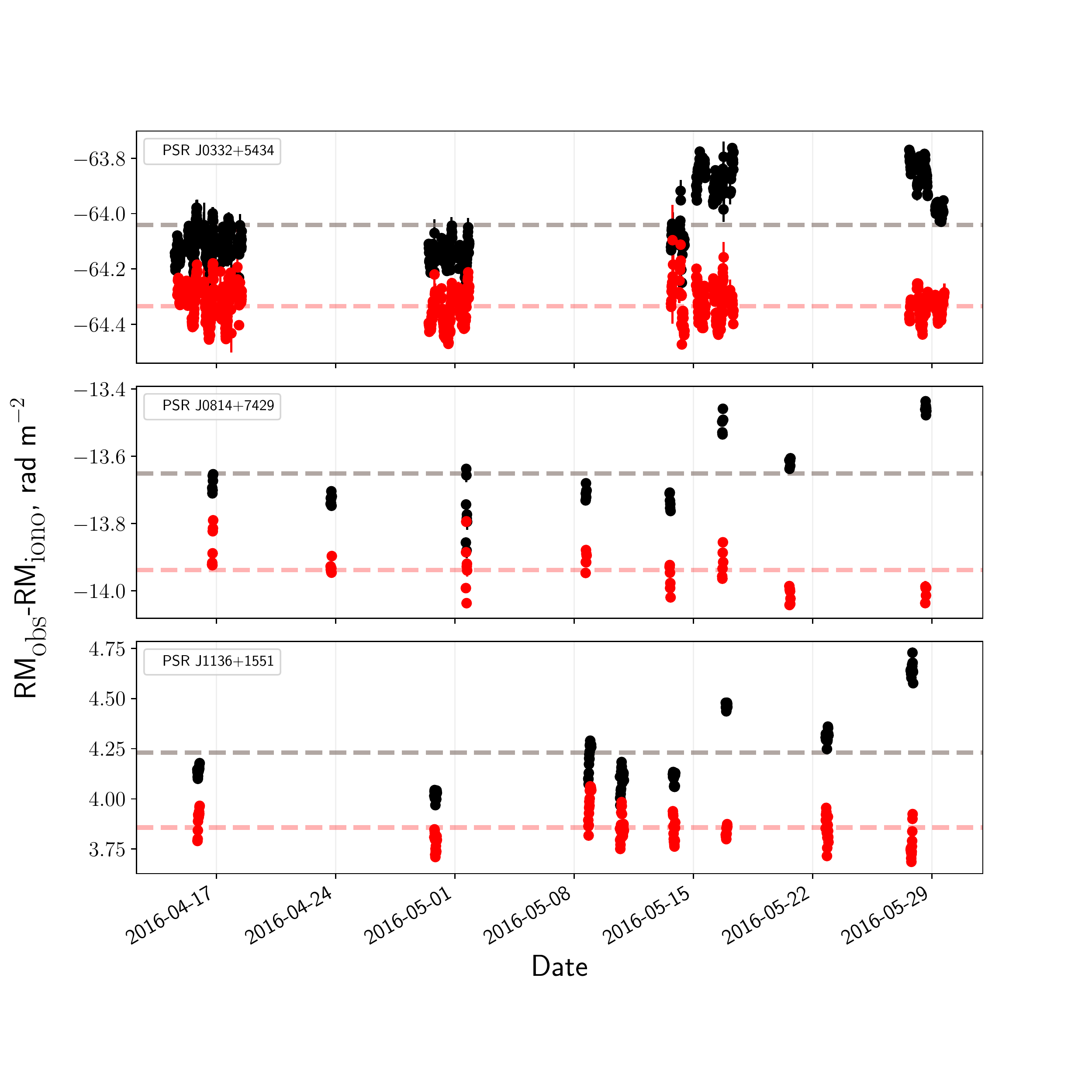}
    \caption{(Colours online) Residuals ($\text{RM}_{\text{obs}} - \text{RM}_{\text{mod}}$), while applying the CODG+POMME10 (black dots) and JPLG+POMME10 (red dots) models for three different pulsars observed with three different GLOW stations. \textit{From top to bottom}: PSRs~ J0332+5434 (with DE609), J0814+7429 (with DE605), J1136+1551 (with DE601).}
    \label{fig:res_three_puls}
\end{figure}

The residual RM series, after subtraction of the ionospheric model from the observed RM values, show the presence of correlated structures and strong colored noise. For instance, Fig.~\ref{fig:res_three_puls} shows the residual RMs for three different pulsars across about a 2-month long timespan. As mentioned in the previous section, the ionospheric RM was corrected by using the RMextract software package, the POMME10 geomagnetic field model, and the CODG/JPLG maps. The CODG time series show similar trend (e.g., the jump of the RM around the 15th of May 2016) despite the pulsars being significantly separated on the sky. The magnitude of these RM variations significantly exceeds those expected from astrophysical sources (see Section \ref{sec:sum} for details). This implies that the origin of the correlations is not interstellar, but an insufficient  modelling of the ionosphere. Moreover, if these maps provide unreliable information about the uncertainties of the TEC values, this will affect the uncertainties of the modelled ionospheric RM and, in turn, our ability to determine the significance of astrophysical RM variations. 

In order to solve for these issues, we conduct an independent search of the systematics in the modelled ionospheric RMs, with the aim of obtaining good estimates of the white-noise level and the uncertainties for the ionospheric RM time series.

\begin{figure}
\includegraphics[width=1.1\columnwidth]{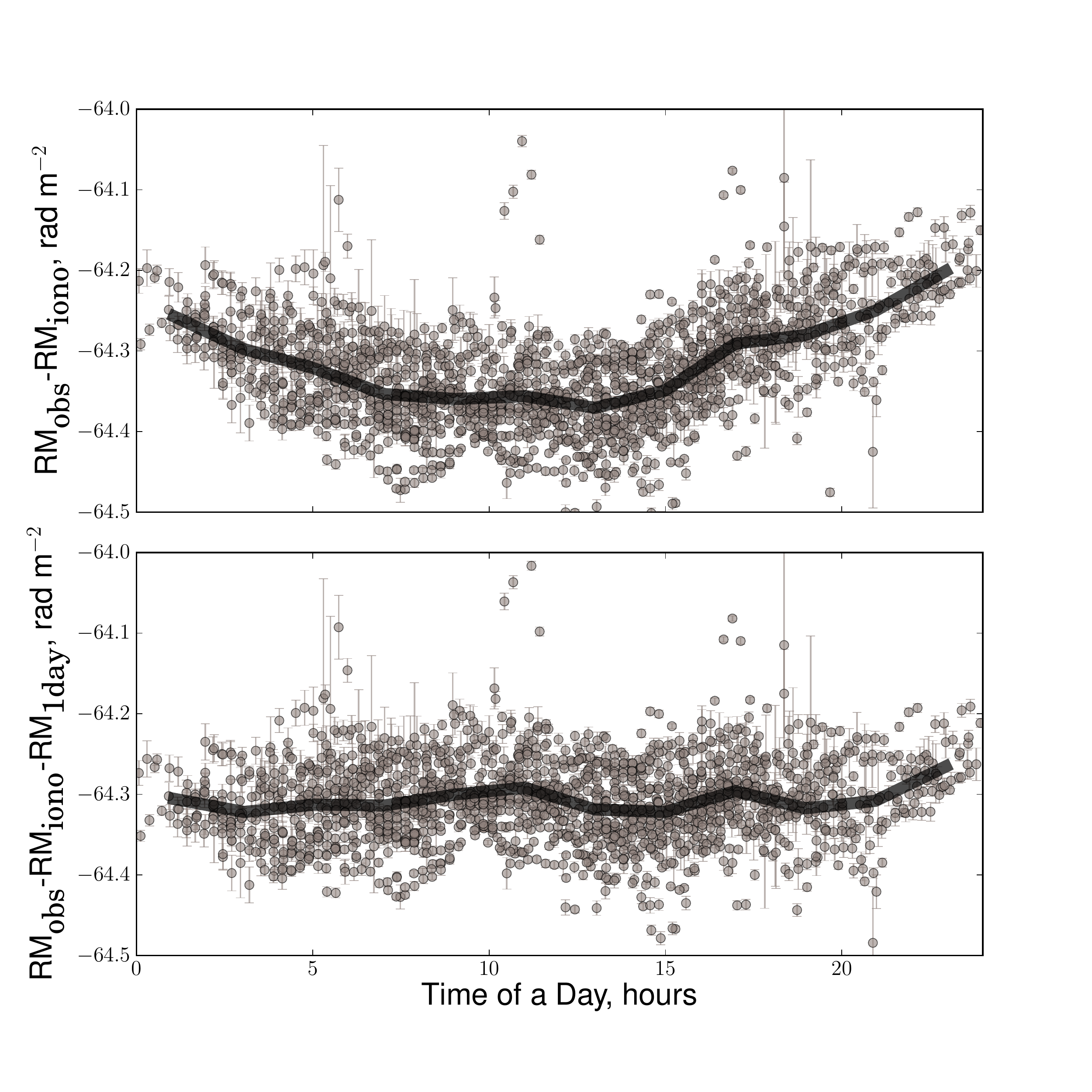}
    \caption{Correlation pattern of the RM residuals after correcting for the ionosphere with respect to the time in a day of observations for PSR~J0332+5434, while using JPLG+POMME10 model. \textit{Upper panel} -- before the subtraction of a 1-day sinusoid. \textit{Lower panel} -- after the subtraction of a 1-day sinusoid. The black thick line on both plots shows the result of data smoothing. }
    \label{fig:time_of_day_figure}
\end{figure}

\begin{figure}
\includegraphics[width=1.1\columnwidth]{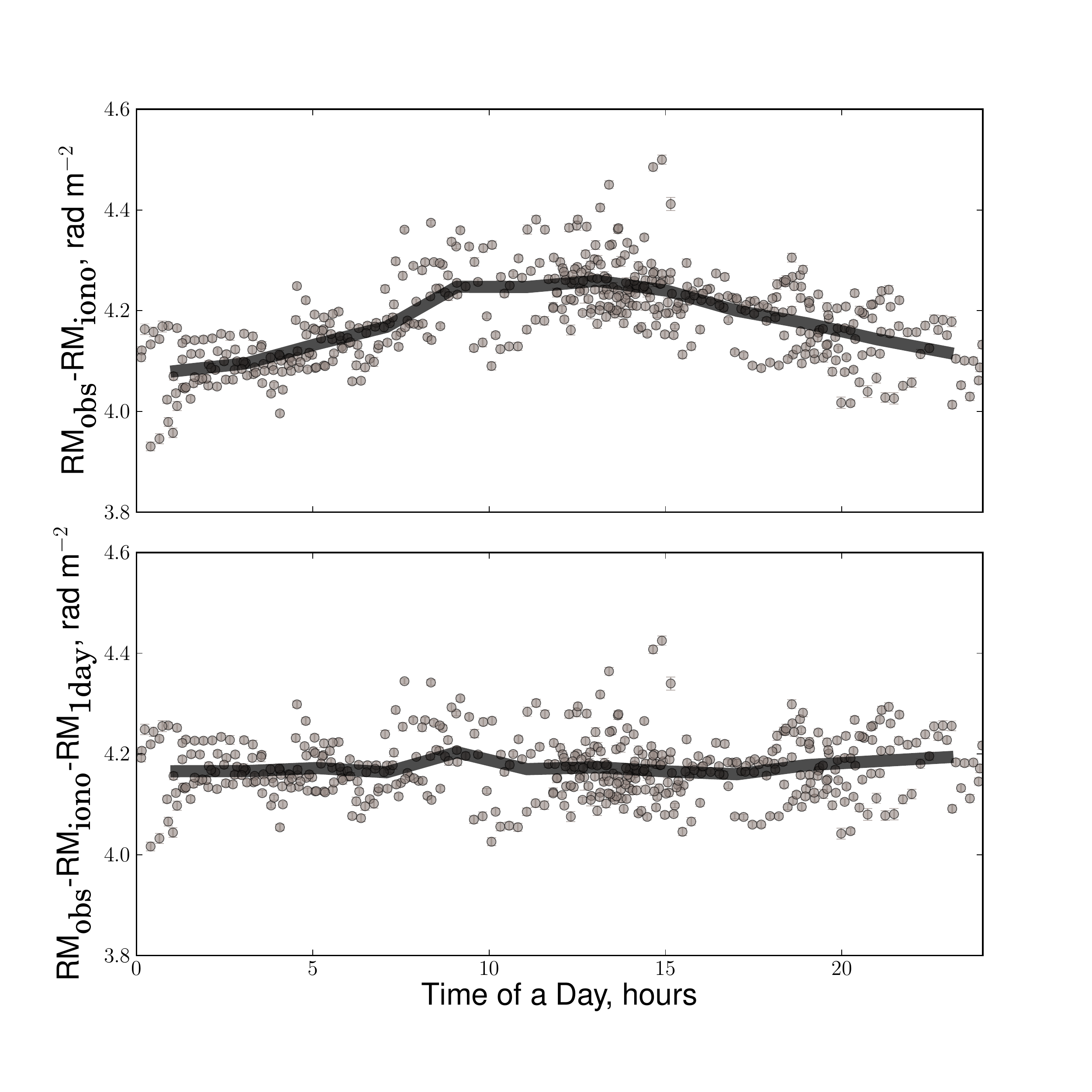}
    \caption{The same as in Fig. \ref{fig:time_of_day_figure} for PSR~J1136+1551 with UQRG+POMME10 model.}
    \label{fig:time_of_day_figure1136}
\end{figure}

Some of the observed structures in the residuals can be well explained by a diurnal sinusoid with amplitude $A_{\text{d}}$, the effect of which is demonstrated in Fig.~\ref{fig:time_of_day_figure}, \ref{fig:time_of_day_figure1136} and Fig. \ref{fig:variat} B,C. Besides being responsible for a 1-day peak in the power spectrum of the RM residuals, it also creates a 1-year pseudo-periodicity in the data, as the transit time of the source shifts gradually during the day across the year. 

\begin{figure}
\includegraphics[width=1.1\columnwidth]{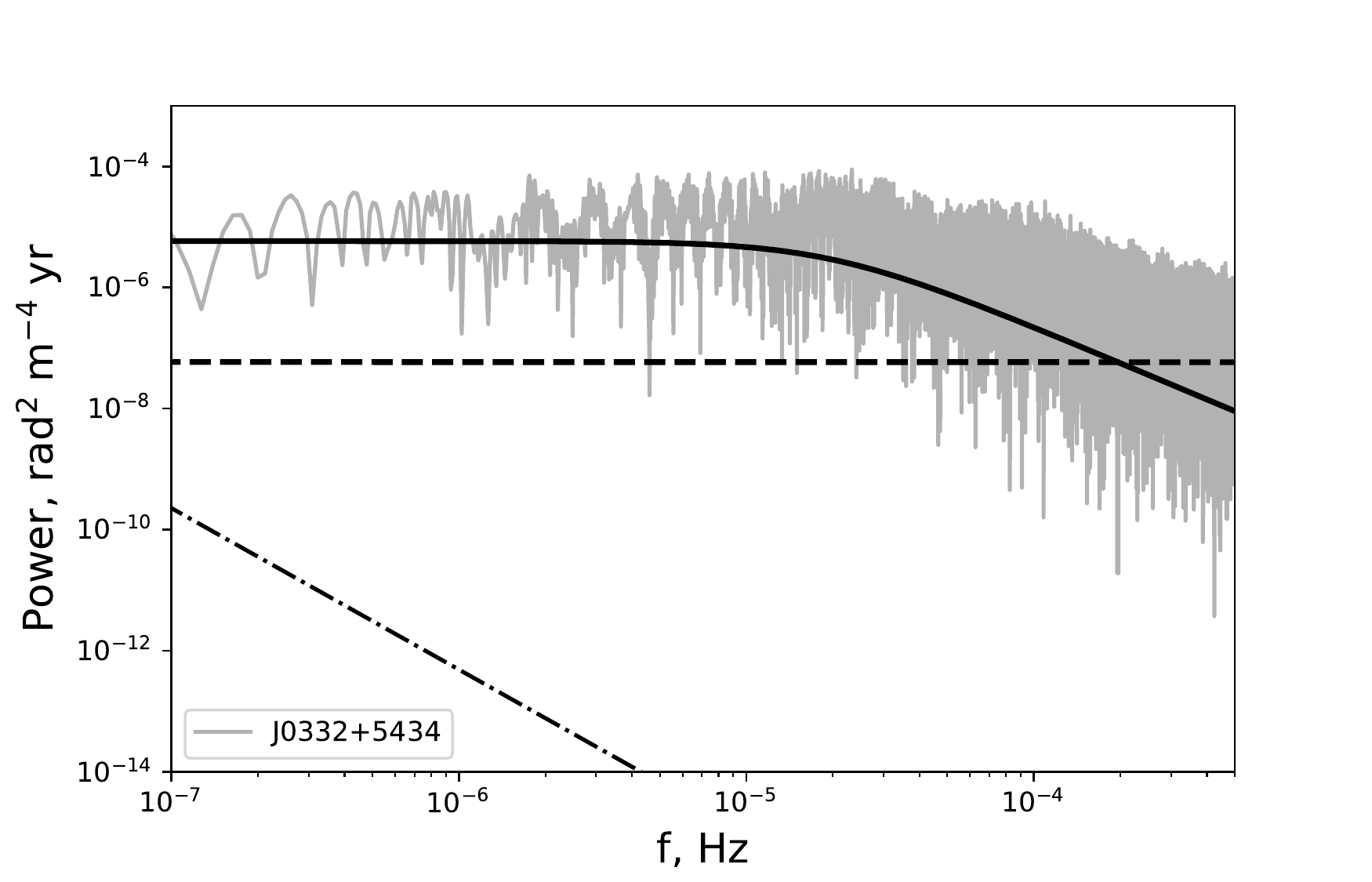}
\includegraphics[width=1.1\columnwidth]{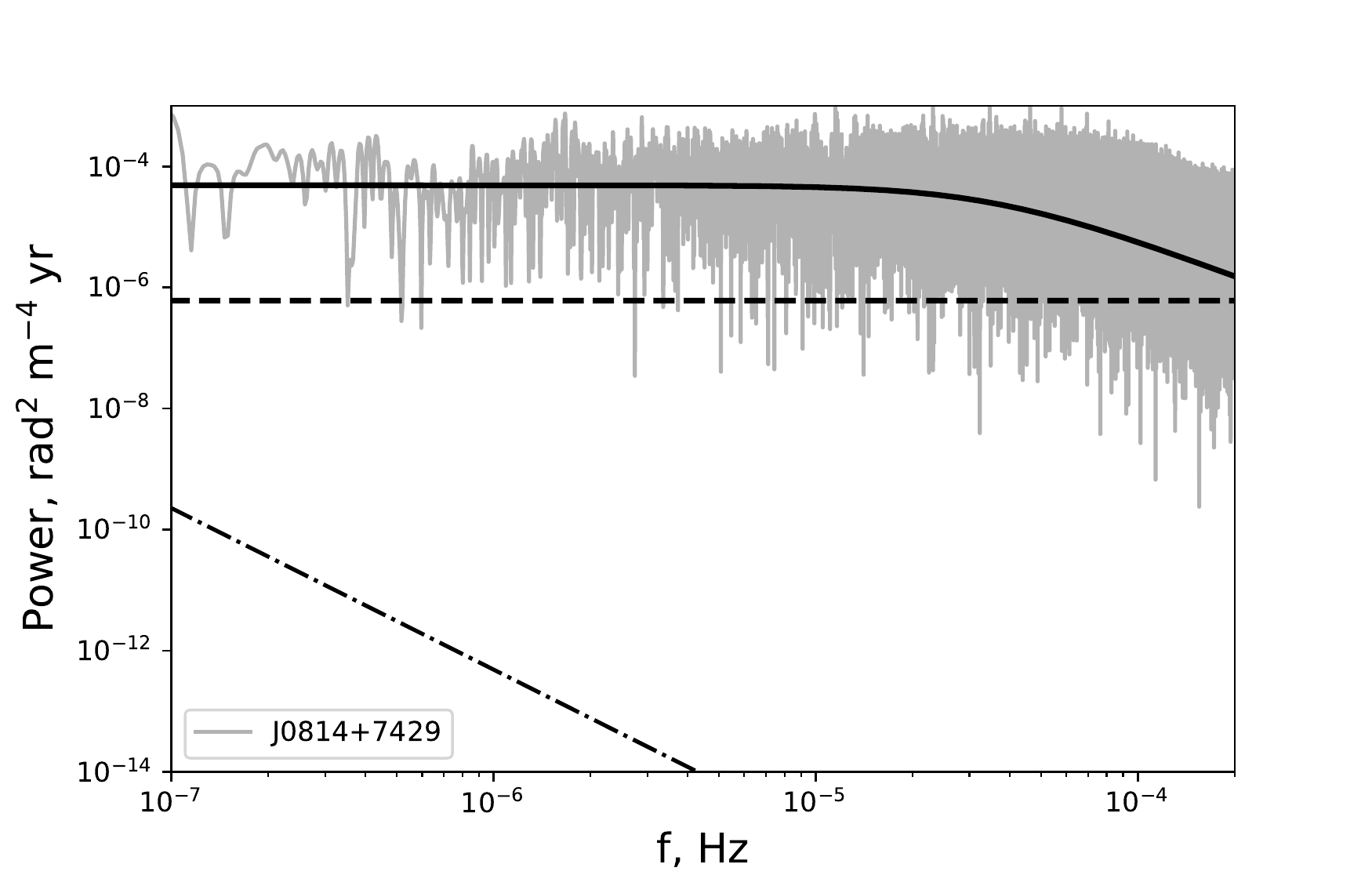}
\includegraphics[width=1.1\columnwidth]{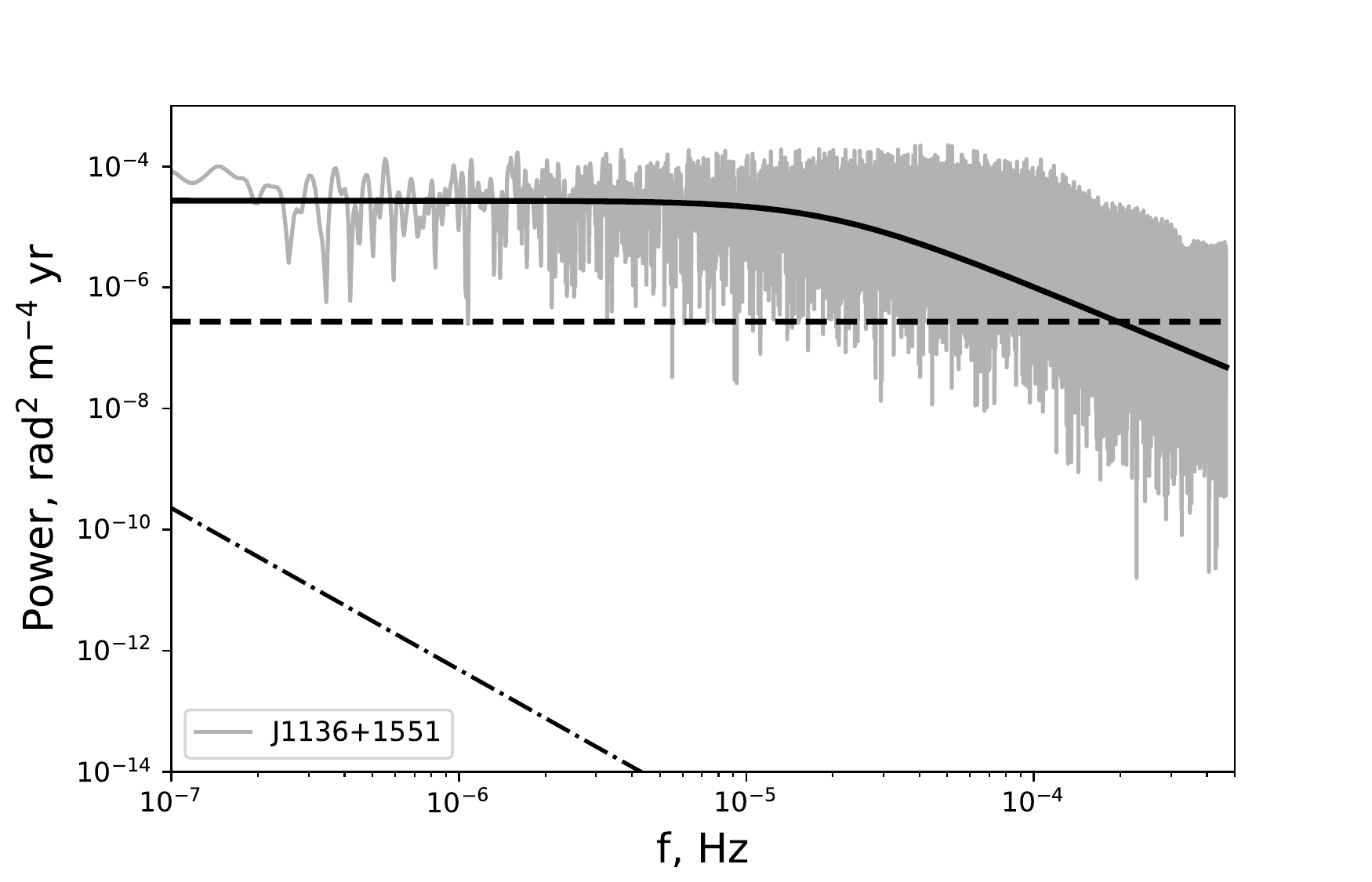}
    \caption{Power spectrum of the residuals ($\text{RM}_{\text{obs}}-\text{RM}_{\text{mod}}-\text{RM}_{\text{1day}}$), shown with grey line, while applying JPLG+POMME10 model to a 6-month datasets of the circumpolar PSRs J0332+5434 (upper plot), J0814+7429 (middle plot) and J1136+1551 (lower plot). The solid black line shows the theoretical shape of a Lorentzian spectrum with $A_{\text{L}}$ and $f_{0}$, defined in Table \ref{tab:example_table1}. The thick dashed line shows the level of the uncorrelated noise, as given by RM measurement uncertainties. The thin dot-dashed line shows the theoretically predicted power spectrum of ionised ISM turbulence (see Eq. (\ref{RM-ism_pred})).}
    \label{fig:spect}
\end{figure}

After subtracting the 1-day sinusoid, the spectrum shows obvious evidence of red noise at high frequencies, and evolves into a white noise plateau at low frequencies (see Fig.~\ref{fig:spect}). Such a spectrum is described in our model by a Lorentzian function, also known as an Ornstein-Uhlenbeck process \citep{1930PhRv...36..823U}. In the next two paragraphs we provide the reader with the mathematical description of the found systematics and introduce the criteria for the comparison of different ionospheric maps.  

\subsection{Analysis of RM residuals on timescales up to one year}
\label{sec:below1yr}

The observational evidences, discussed in Section~\ref{sec:system_in_shortdata} allow us to define a mathematical model to describe the contributions to the observed RM time series with timescale shorter than a year. As will be pointed out in Section \ref{sec:sum}, the interstellar contribution $\text{RM}_{\text{IISM}}$ is very small, and typically only visible on timescales of order of several years \citep{2011Ap&SS.335..485Y}. We have restricted the dataset considered in this section to only several months (Table \ref{table1}), so we can assume this parameter to be constant and, thus, the contribution $\text{RM}_{\text{IISM}}$ will not bias the estimates of the parameters of the systematics.

The vector $\textbf{RM}_{\text{obs}} = [\text{RM}_{\text{t}_1}, \text{RM}_{\text{t}_2},..., \text{RM}_{\text{t}_N}]$ that contains the RM time series of a certain pulsar observed at $N$ epochs $t_i$ can be seen as a combination of deterministic and stochastic contributions:
\begin{equation}
\textbf{RM}_{\text{obs}}=\underbrace{\textbf{RM}_{\text{iono}}+\textbf{RM}_{\text{1day}}+\text{RM}_{\text{IISM}}}_{\text{deterministic}}+\underbrace{\textbf{RM}_{\text{noise}} + \bm{n}}_{\text{stochastic}}.
\label{modelion}
\end{equation}
$\textbf{RM}_{\text{iono}}$ stands for the semi-empirical ionospheric thin layer model of RM variations described in Section \ref{sec:model_ion}. $\textbf{RM}_{\text{1day}}$ is the harmonic signal with 1-day period, that can be parametrised as $\textbf{RM}_{\text{1day}}=A_{\text{d}} \sin (2\pi\bm{t}/1\text{day} + \phi)$. $\textbf{RM}_{\text{noise}}$ and $\bm{n}$ are stochastic noise contributions. $\textbf{RM}_{\text{noise}}$ is given by the plateau of the Lorentzian spectrum, whose one-sided spectral density is described as:
\begin{equation}
S(f)=\frac{A_{\text{L}}^2}{f_0\left[1+\left(\frac{f}{f_0}\right)^2\right]},
\end{equation}
with $A_\text{L}$ [rad m$^{-2}$] being the already mentioned amplitude of the stochastic signal and $f_0$ [day$^{-1}$] the turnover frequency. From this expression it can be easily shown that, while behaving like red noise on a short time scales, $\textbf{RM}_{\text{noise}}$ reduces to white noise for $f\ll f_0$ with a constant variance $A_\text{L}^2$. By applying the Wiener-Khinchin theorem, the variance-covariance matrix of this process is then given by:
\begin{equation}
C_{\text{L}}=A_\text{L}^2 \exp(-f_{0} \tau),
\end{equation}
where $\tau=2\pi|t_i-t_j|$ with $t_i$ and $t_j$ are two different epochs.

The uncorrelated white noise component $\bm{n}$ in Eq.~(\ref{modelion}), coming from the measurement noise of Stokes parameters (see Fig. \ref{pic:gist}), has a flat power spectral density with variance-covariance of the form:
\begin{equation}
\label{cwn}
C_{\text{WN}}=\sigma^2_i \delta_{ij},
\end{equation}
with $\delta_{ij}$ being a Kronecker delta and $\bm{\sigma}$ the vector of the formal uncertainties of the observed RMs\footnote{The uncertainties are modified by a factor $\eta$ ,see Eq.~(\ref{eta})}, determined via the Bayesian Lomb-Scargle Periodogram described at the end of Appendix~\ref{app:rm}. 

In order to investigate the properties of the stochastic and deterministic signals that emerge in the RM residuals after the ionospheric correction, we use Bayesian inference in the time domain. Given the model in Eq.~(\ref{modelion}), and assuming that the stochastic parts are drawn from random Gaussian processes, the posterior probability for the unknown parameters $\mathbf{\Theta}=[A_\text{L}, f_{0}, A_\text{d}, \phi, \text{RM}_{\text{IISM}}]$ is written as:
\begin{align}
\begin{split}
\log P_{\text{pst}}(\mathbf{\Theta}) \sim \log P_{\text{pr}}(\mathbf{\Theta})\\
-\sum_{i=1}^{N} \frac{1}{2}\left(\textbf{RM}_{\text{obs}}-\textbf{RM}_{\text{iono}}-\textbf{RM}_{\text{1day}}-\text{RM}_{\text{IISM}}\right) \times\\
C^{-1}\times\left(\textbf{RM}_{\text{obs}}-\textbf{RM}_{\text{iono}}-\textbf{RM}_{\text{1day}}-\text{RM}_{\text{IISM}}\right)
-\frac{1}{2}\ln(2\pi \text{det} C),
\label{eq:post}
\end{split}
\end{align}
where $C=C_{\text{L}}+C_{\text{WN}}$.

We have applied the model discussed in this paragraph to pulsar datasets that span less than a year (outlined in Table~\ref{table1}). 
The high S/Ns of our pulsars make us more sensitive to the signals generated by the imperfections of ionospheric RM modelling, described by Eq.~(\ref{modellin}). The factor $f_{B}$ was fixed to 1.11, as found in Section \ref{subsec:linear}, and, thus, was excluded from the set of free parameters of the model.

\begin{figure*}
\includegraphics[width=1.1\textwidth]{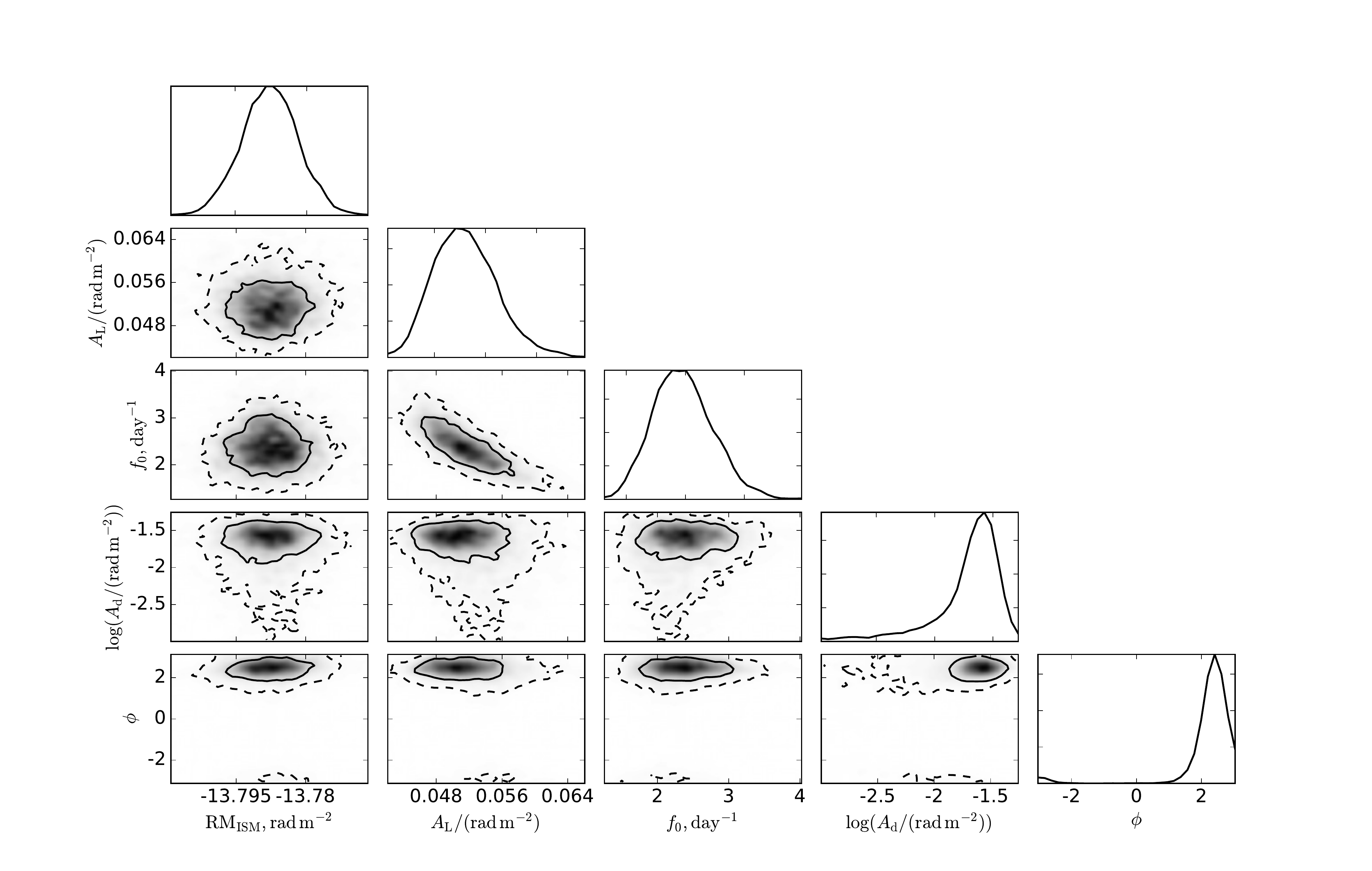}
    \caption{One and two-dimensional posterior distribution for a subset of the noise parameters that characterize the RM residuals of PSR~J0814+7429 after the subtraction of the ionospheric model (using JPLG maps+POMME10 geomagnetic model). \textit{From left to right} -- rotation measure of ionised ISM $\text{RM}_{\text{IISM}}$ [rad m$^{-2}$], which is assumed to be constant on time scales of several months, the level of the white noise plateau $A_\text{L}$ [rad m$^{-2}$] in the Lorentzian spectrum, the turnover frequency $f_0$ [day$^{-1}$] of the Lorentzian spectrum, the amplitude of the 1-year harmonic signal in the residuals $A_\text{d}$ [rad m$^{-2}$], the phase of the harmonic signal $\phi$.}
    \label{fig:marg_figure}
\end{figure*}

In order to explore the 5-dimensional parameter space $\mathbf{\Theta}$ of Eq.~(\ref{modelion}), we ran a Markov Chain Monte Carlo simulation, using the Bayesian inference tool \textsc{MultiNest} \citep{2009MNRAS.398.1601F}. The priors of the parameters $P_{\text{pr}}(\Theta)$ were chosen to be uninformative \citep[see][]{2016MNRAS.457.4421C}: uniform for $A_{\text{L}}$, $f_{0}$, $\phi$, and $\text{RM}_{\text{IISM}}$ and log-uniform for $A_\text{d}$. A representative example of the obtained 2-dimensional posterior probability plot is shown in Fig.~\ref{fig:marg_figure}.

Because we only used data from a limited group of pulsars, and probed a statistically not significant sample of LoS, the values given in Table \ref{tab:example_table1} should not be treated as definitive solutions. However, our results give a qualitative estimation of the accuracy of different ionospheric maps.

\subsection{Analyis of RM residuals on timescales beyond one year}
\label{subsec:linear}

\begin{figure}
\includegraphics[width=1.1\columnwidth]{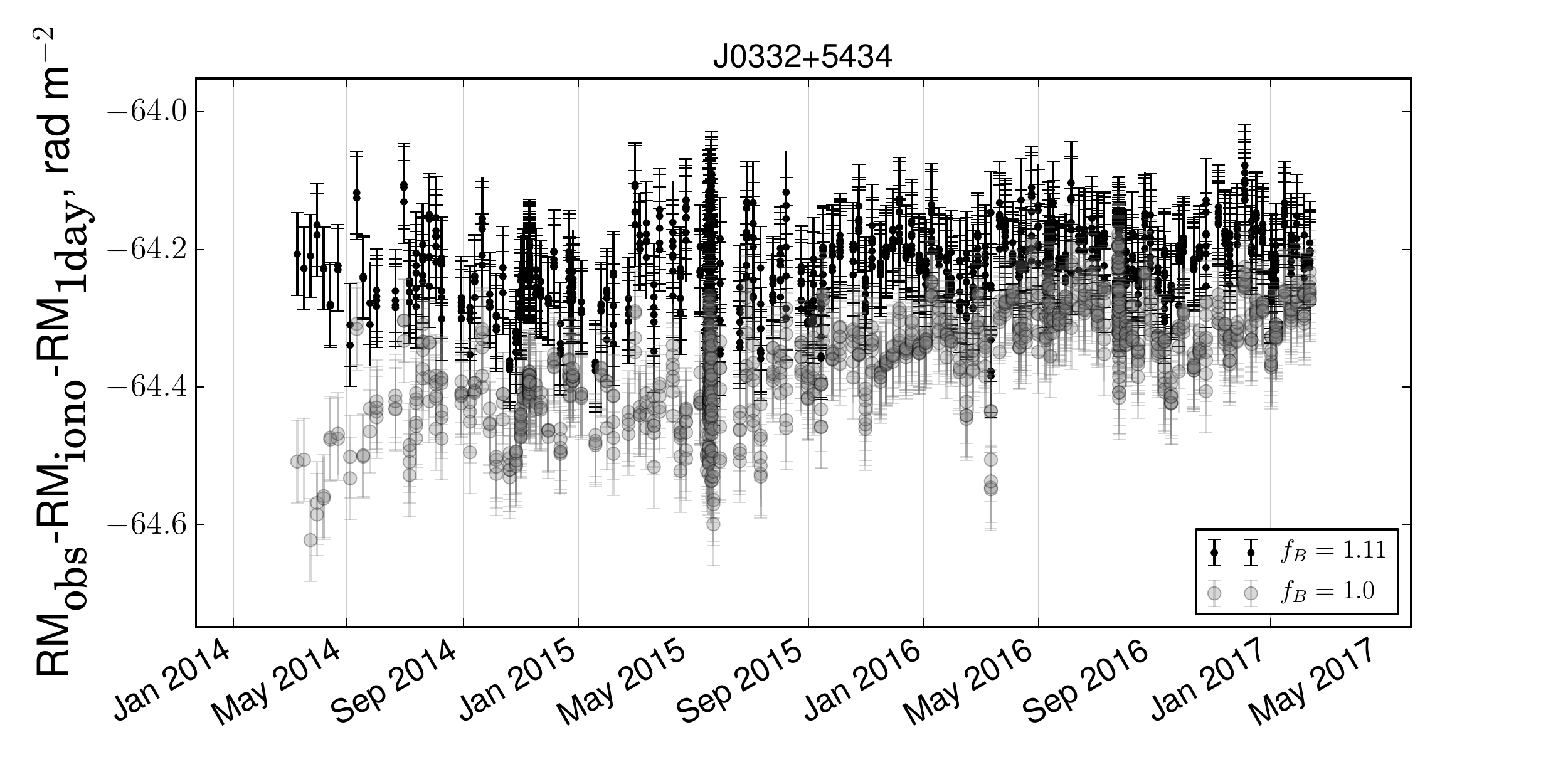}
\includegraphics[width=1.1\columnwidth]{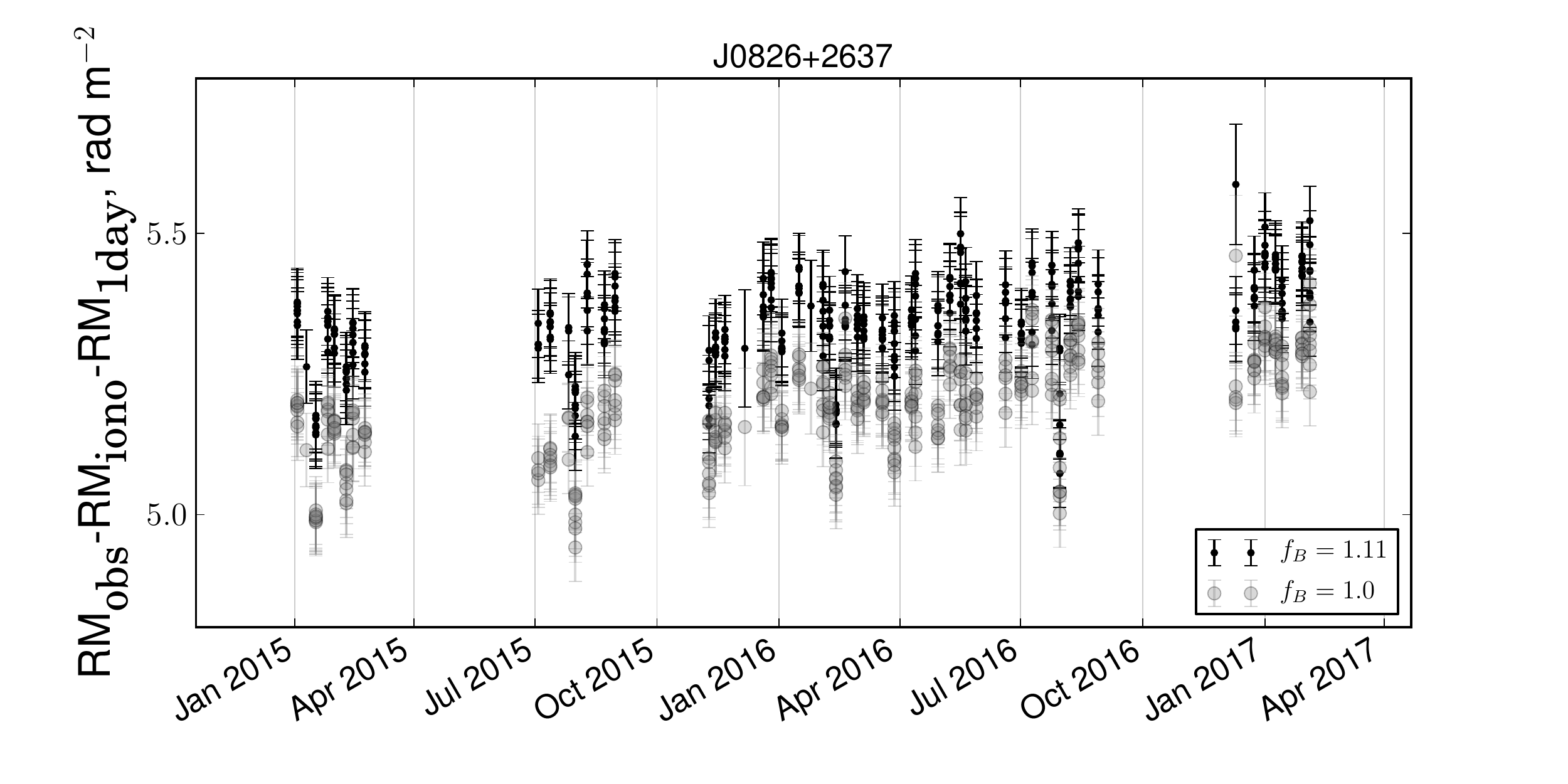}
\includegraphics[width=1.1\columnwidth]{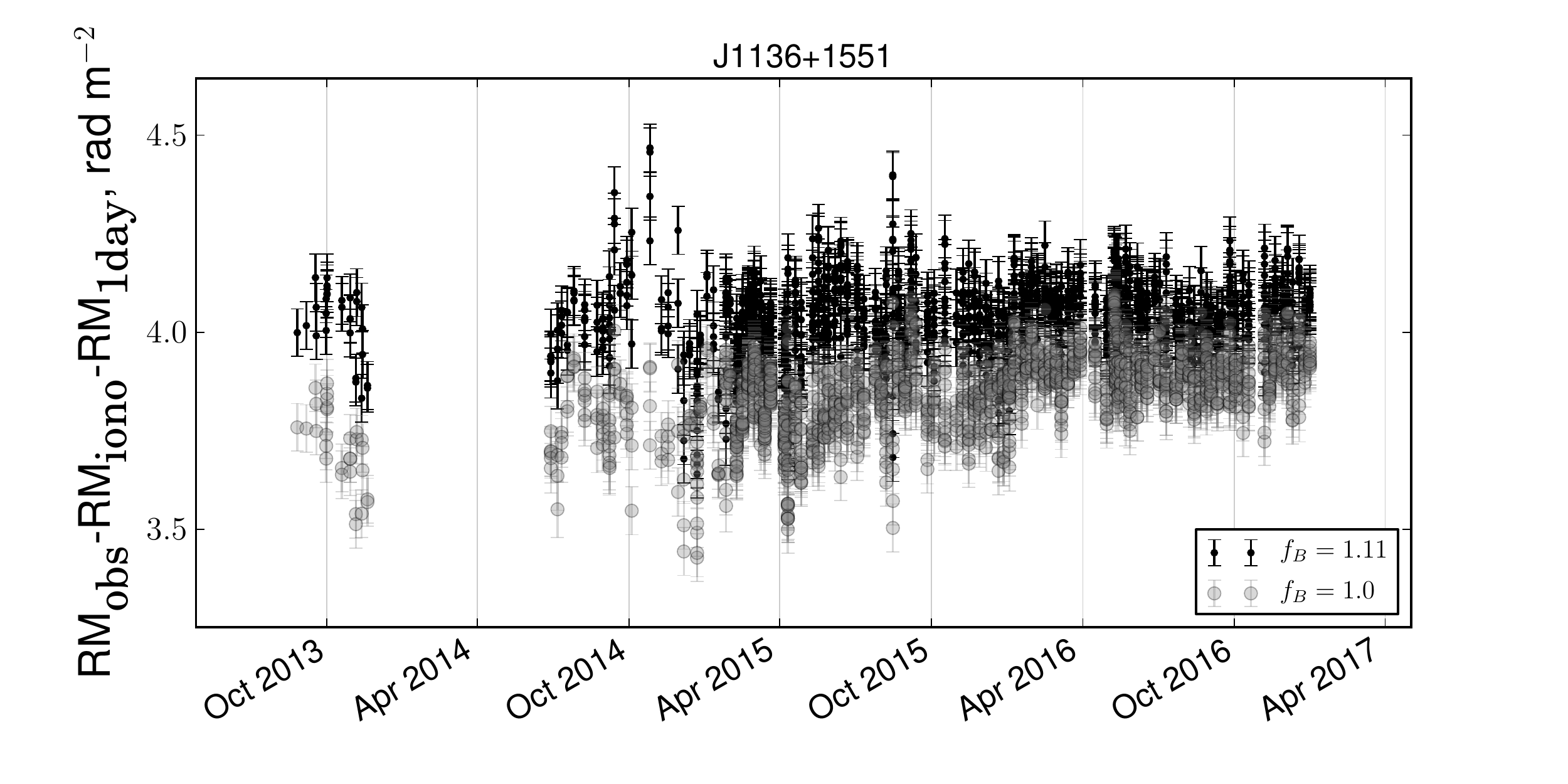}
\includegraphics[width=1.1\columnwidth]{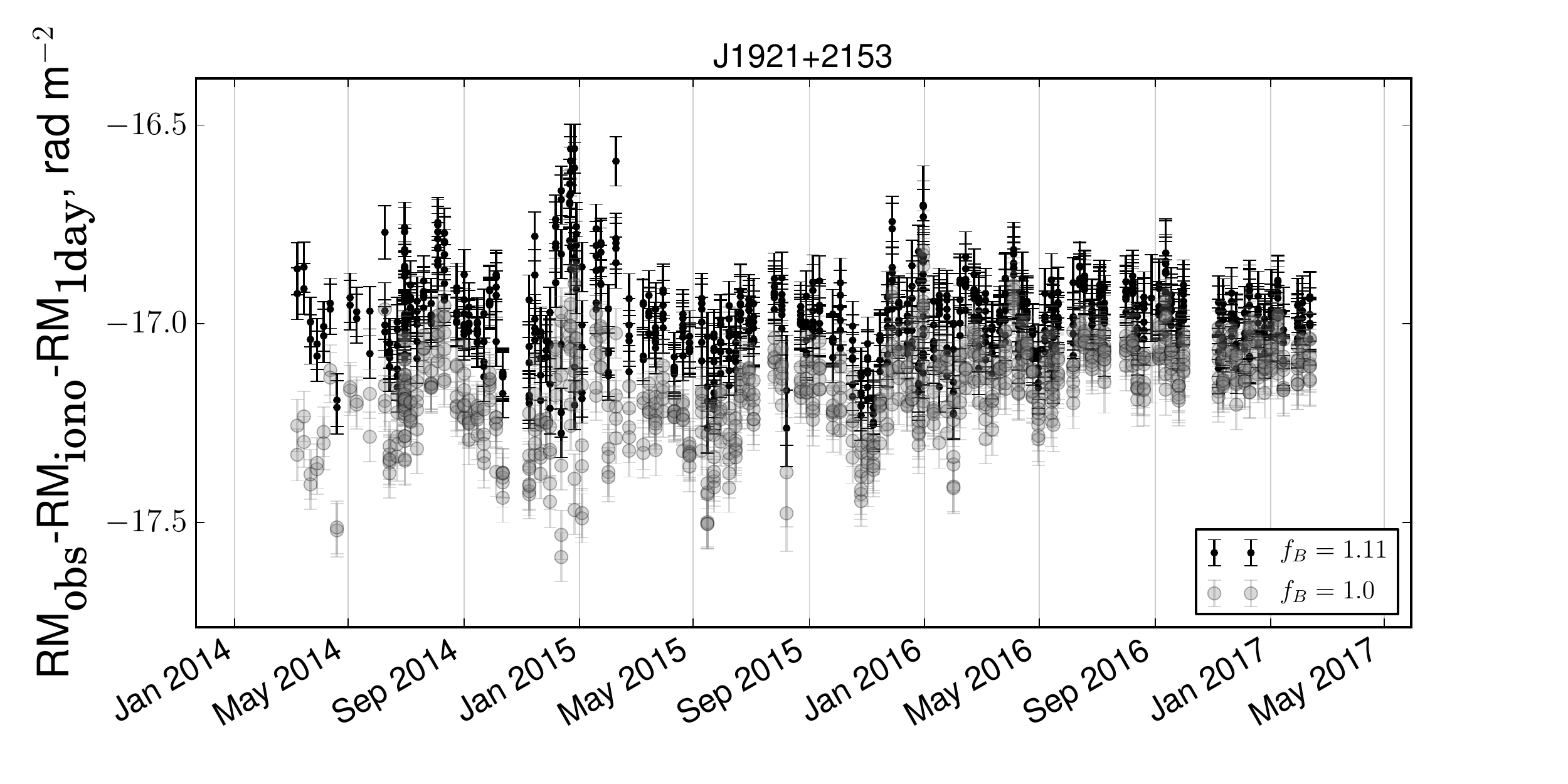}
    \caption{Absorption of the linear trend due to the application of the $f_B$ factor. The grey circles correspond to the $f_B=1$, black dots to $f_B=1.11$. We here use three years of data for (from top to bottom) PSRs~J0332+5434, J0826+2637, J1136+1551, J1921+2153. The ionospheric contribution is modelled with JPLG maps combined with POMME10 geomagnetic model.}
    \label{fig:lin_trend}
\end{figure}

\begin{figure}
\includegraphics[width=1.1\columnwidth]{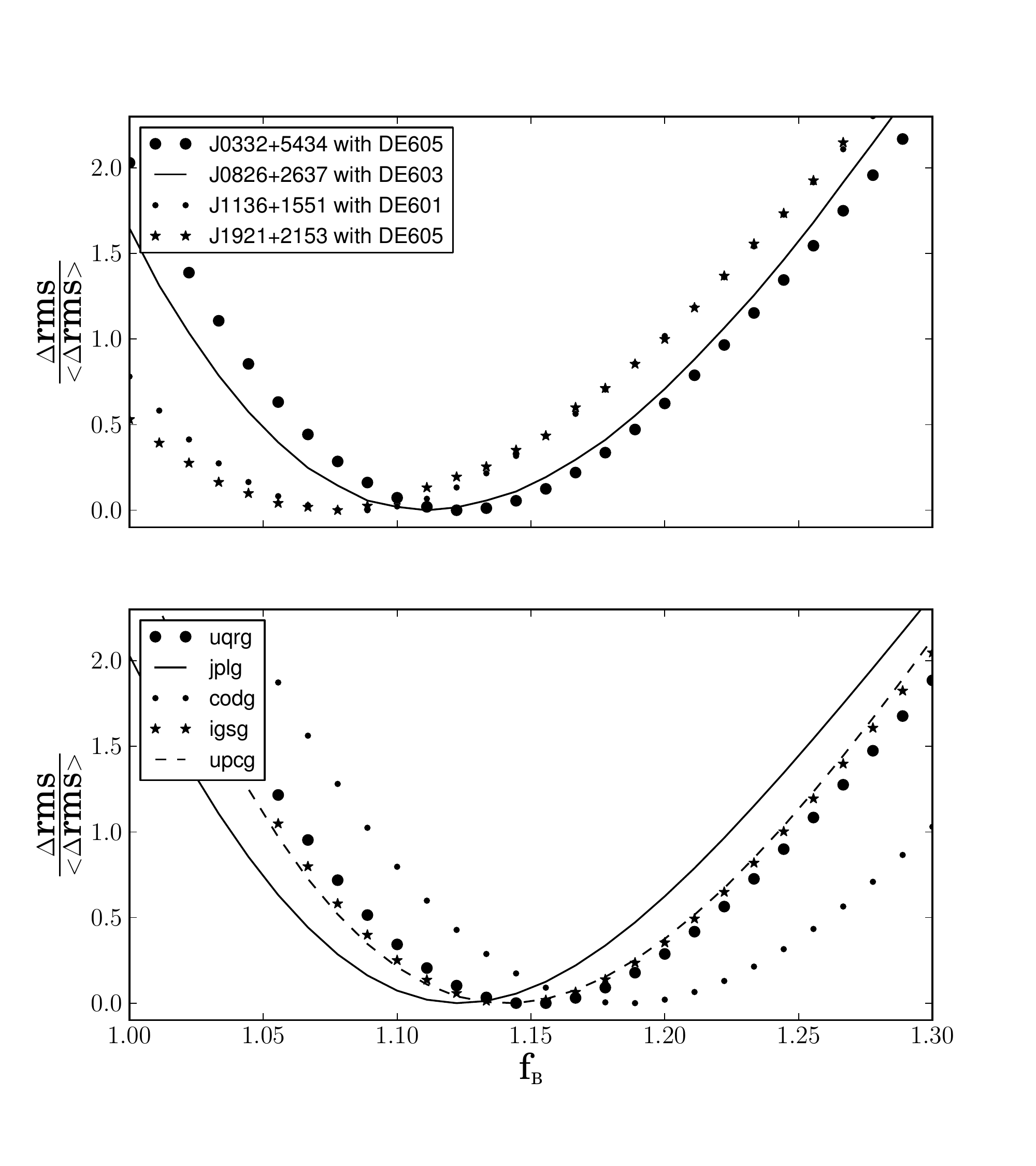}
    \caption{\textit{Upper panel}: Rms of the RM residuals obtained by using the JPLG map (and normalised with respect to the minimum value for each case), vs the $f_B$ factor. We here use 3 years of data for PSRs~J0332+5434, J0826+2637, J1136+1551, J1921+2153 (see Table \ref{table1}). \textit{Lower panel}: Rms of the RM residuals obtained by using UQRG, JPLG, CODG, IGSG, and UPCG maps (and normalised with respect to the minimum value for each case) vs the $f_B$ factor. We here use 3 years of data for PSR~J0332+5434 observed with DE605 (see the text for more details). The trends show clear improvements of the modelling when using $f_B \simeq 1.10$ -- $1.14$}
    \label{fig:3p_min}
\end{figure}

After subtracting $\textbf{RM}_{\text{iono}}$ and $\textbf{RM}_{\text{1day}}$ from the observed RM time series, the long term datasets still show a deterministic linear trends. The trend is not visible on a timescale of months, but it becomes obvious across several years (see Fig.~\ref{fig:lin_trend}). We find that such linear trend can be suppressed by scaling the $\textbf{RM}_{\text{iono}}$ time series (resulting from a thin layer model) by a constant factor $f_{B}$. In other words, Eq.~(\ref{modelion}) is modified as
\begin{equation}
\label{modellin}
\textbf{RM}_{\text{obs}}=\frac{\textbf{RM}_{\text{iono}}}{f_{B}}+\textbf{RM}_{\text{1day}}+\text{RM}_{\text{IISM}}+\textbf{RM}_{\text{noise}} + \bm{n}.
\end{equation}
A positive trend of the order of $1\text{--}2\times 10^{-4}$ rad m$^{-2}$ day$^{-1}$ was noticed in four pulsars observed with three GLOW stations. Removal of the linear trend, by applying the factor $f_{B}$, is demonstrated in Fig. \ref{fig:lin_trend}. 

Making use of pulsar datasets that span more than one year (outlined in Table \ref{table1}) and all six global ionospheric maps, considered in this paper, the least square fit estimate was found to be $f_{B}=1.11^{+0.04}_{-0.04}$.  The decrease in root-mean-square (rms), mostly due to the elimination of the linear trend, is illustrated in Fig. \ref{fig:3p_min}. The results for all three geomagnetic models are identical.

There are several physical interpretations possible for the factor $f_B$. Among them is the possible overestimation of the geomagnetic strength, $B_{\text{LoS}}$, and/or the underestimation of the ionospheric effective height \citep{2002RaSc...37.1015B} due to the poor knowledge about the electron density in Earth's plasmasphere. For instance, $f_{B}=1.11$ is equivalent to an increase of the effective height from 450 km up to $\sim$ 700 km. An explanation to the trend might be searched in the complex dynamical behavior of the ionospheric effective height. As a matter of fact, it has been shown in multiple investigations \citep[e.g.][]{2011JGeod..85..887H, 2016PASA...33...31A}, that the ionospheric effective height can vary from 300 up to 800 km, depending on the time of day, season, and level of Solar activity. For instance, the 11-year sunspot cycle, the last maximum of which was in 2014, can cause significant ionization in the ionospheric layer, thus both increasing the ionospheric thickness and ionospheric effective  height \citep{2007JGRA..11211311L}. One promising way to improve the model is by using the effective heights determined via the IRI-Plas software \citep{2013JASTP.102..329G}, which takes into account plasmasphere contribution \citep{2016PASA...33...31A}.

The nature of the factor $f_{B}$ is still under investigation and is planned to be tested on a larger sample of pulsars in the future.

\begin{table*}
	\centering
	\caption{ Estimation of the noise parameters based on the Bayesian analysis of RM residuals using POMME10 geomagnetic model and different ionospheric maps. The results for other two considered in this paper geomagnetic models (EMM and IGRF) are indistinguishable within the uncertainties and are demonstrated in Supplementary material online. The analysis was based on datasets of PSR~J0332+5434,  PSR~J1136+1551 and PSR~J0814+7429, respectively (see Table \ref{table1}). The used noise model is the one described in Eq.~(\ref{modellin}). We report the median and 1-sigma uncertainty values of the amplitude of the Lorentzian spectrum $A_\text{L}$ [rad m$^{-2}$], which effectively quantifies the measurement uncertainties of the ionospheric RM corrections; the turnover frequency of the Lorentzian spectrum $f_0$ [day$^{-1}$], the amplitude of the 1-day harmonic signal $A_\text{d}$ [rad m$^{-2}$] and maximum likelihood estimation of $\text{RM}_{\text{IISM}}$ [rad m$^{-2}$]. The latter is assumed to be constant across a timescale of several months. The factor $f_{B}=1.11$ was applied.}
	\label{tab:example_table1}
	\begin{tabular}{lccccccr} 
		\hline
        &&PSR J0332+5434\\
        \hline
		Model & $A_\text{L}^{\text{med}}$ & $f_0^{\text{med}}$ & $A_\text{d}^{\text{med}}$ & $\text{RM}_{\text{IISM}}^{\text{ML}}$ \\
		\hline
        \hline
		UQRG &  $0.045_{-0.002}^{+0.003}$ &  $1.5_{-0.2}^{+0.2}$ & $0.012_{-0.006}^{+0.007}$ & $-$64.16 \\
        \\
		JPLG &  $0.050_{-0.002}^{+0.002}$ &  $1.7_{-0.2}^{+0.2}$ & $0.025_{-0.005}^{+0.007}$ & $-$64.21\\
	        \\
		EHRG & $0.054_{-0.003}^{+0.003}$ &  $1.2_{-0.1}^{+0.1}$ &  $0.012_{-0.008}^{+0.007}$ & $-$64.05\\
         \\
          IGSG & $0.060_{-0.003}^{+0.003}$ &  $1.2_{-0.1}^{+0.1}$ & $0.02_{-0.007}^{+0.005}$ & $-$64.08\\
        \\
         ESAG & $0.068_{-0.004}^{+0.005}$  & $1.1_{-0.1}^{+0.1}$ & $0.025_{-0.009}^{+0.007}$ & $-$64.05\\
         \\
        UPCG & $0.073_{-0.004}^{+0.003}$ &  $0.9_{-0.1}^{+0.1}$ & $0.025_{-0.014}^{+0.008}$ & $-$64.17 \\
        \\
        CODG & $0.12_{-0.01}^{+0.01}$ & $0.29_{-0.06}^{+0.06}$ & $0.063_{-0.009}^{+0.009}$ & $-$63.95\\
        \hline
	\end{tabular}
    \\
     \begin{tabular}{lccccccr} 
		\hline
        &&PSR J1136+1551\\
        \hline
		Model & $A_\text{L}^{\text{med}}$ & $f_0^{\text{med}}$ & $A_\text{d}^{\text{med}}$ & $\text{RM}_{\text{IISM}}^{\text{ML}}$ \\
		\hline
        \hline
		UQRG & $0.061_{-0.004}^{+0.005}$ & $2.1_{-0.3}^{+0.3}$ & $0.079_{-0.009}^{+0.01}$ &  4.16 \\
        \\
		JPLG & $0.073_{-0.004}^{+0.004}$ & $1.9_{-0.5}^{+0.6}$ & $-$ & 4.02 \\
        \\
        EHRG & $0.082_{-0.004}^{+0.005}$ & $1.3_{-0.2}^{+0.2}$ & $0.03_{-0.02}^{+0.01}$ & 4.22 \\
        \\
        IGSG & $0.142_{-0.01}^{+0.008}$ & $0.6_{-0.1}^{+0.1}$ & $0.04_{-0.02}^{+0.02}$ & 4.19 \\
        \\
        ESAG & $0.110_{-0.007}^{+0.008}$ & $1.0_{-0.1}^{+0.1}$ & $0.03_{-0.01}^{+0.02}$ & 4.26 \\
        \\
        UPCG & $0.123_{-0.008}^{+0.010}$ & $0.9_{-0.1}^{+0.2}$  & $0.08_{-0.02}^{+0.02}$ & 4.18 \\
        \\
        CODG & $0.21_{-0.02}^{+0.02}$ & $0.14_{-0.03}^{+0.03}$ &   $0.08_{-0.02}^{+0.02}$  & 4.18\\
		\hline
	\end{tabular}
    \\
	\begin{tabular}{lccccccr} 
		\hline
        &&PSR J0814+7429\\
        \hline
		Model  & $A_\text{L}^{\text{med}}$ & $f_0^{\text{med}}$ &  $A_\text{d}^{\text{med}}$ & $\text{RM}_{\text{IISM}}^{\text{ML}}$ \\
		\hline
        \hline
        UQRG & $0.053_{-0.003}^{+0.004}$ & $2.7_{-0.5}^{+0.5}$ & $0.049_{-0.007}^{+0.006}$ &  $-$13.75 \\
        \\
        JPLG & $0.051_{-0.003}^{+0.004}$ &$2.3_{-0.4}^{+0.4}$ &   $0.024_{-0.008}^{+0.006}$ & $-$13.79 \\
        \\
        EHRG & $0.054_{-0.003}^{+0.003}$ & $2.1_{-0.3}^{+0.3}$ & $0.033_{-0.02}^{+0.01}$ & $-$13.66 \\
        \\
        IGSG & $0.064_{-0.004}^{+0.005}$ & $1.4_{-0.3}^{+0.3}$ &  $0.047_{-0.008}^{+0.01}$ & $-$13.69 \\
        \\
        ESAG & $0.067_{-0.005}^{+0.005}$  & $1.5_{-0.3}^{+0.3}$ &  $0.03_{-0.01}^{+0.01}$ & $-$13.65 \\
        \\
         UPCG & $0.069_{-0.005}^{+0.005}$  & $1.4_{-0.3}^{+0.3}$ &  $0.045_{-0.01}^{+0.01}$ & $-$13.74 \\
        \\
        CODG & $0.10_{-0.01}^{+0.01}$ & $0.7_{-0.1}^{+0.1}$ &  $0.07_{-0.01}^{+0.02}$  & $-$13.62\\
		\hline
	\end{tabular}
\end{table*}

\section{Discussions and Conclusions}
\label{sec:sum}
In this paper, we have characterised and investigated the deterministic and stochastic RM variations generated by the ionospheric layer through pulsar observations taken with the German LOFAR stations. The main day-to-day variability was modelled by assuming a thin-layer ionosphere, located at 450 km above the Earth's surface. For this model, the magnetic field was taken from the publicly available geomagnetic maps (POMME10, EMM, IGRF), while the information about electron densities was extracted from the selection of different global ionospheric maps. Besides that, an additional signal peaked at a frequency of 1 day$^{-1}$ in the power spectrum, which was significant in almost all processed datasets, and was removed by including in the model a 1-day period sinusoid. The residual noise could be described by a Lorentzian spectrum, which behaves like white noise on long timescales and defines our sensitivity to long-term RM variations. The parameters of the model were estimated by applying a Bayesian framework to the RM time series of three pulsars. The observed RM for each epoch was determined by using an improved RM synthesis technique, based on BGLSP, which accounts for non-regularly sampled data and constant offsets in Stokes $Q$ and $U$ due to instrumental effects. By running a Markov Chain Monte Carlo, we have estimated the amplitude of the Lorentzian spectrum (or variance of white noise) for all the ionospheric maps. 
An additional linear trend becomes visible on a timescale of several years. To account for this, we have applied a factor $f_{B}=1.11$ to the ionospheric RM contribution that was modelled by RMextract, $\textbf{RM}_{\text{mod}}$, as determined in Section~\ref{subsec:linear}. This slightly reduces the level of the Lorentzian spectrum plateau for some of the pulsars, determined in Section~\ref{sec:below1yr}.

Our results for the three pulsars are slightly different. Nevertheless, two of them (PSRs J0332+5434, J0814+7429) are consistent within 2 sigma, while J1136+1551 shows slightly higher values. We show that
geomagnetic models mostly agree and that consequently the accuracy of
ionospheric RM corrections is dominated by the uncertainties and
inaccuracies in ionospheric TEC maps, which we have investigated in the paper. On average UQRG and JPLG, combined with one of the geomagnetic models, show better results than the other ionospheric maps. If one is going to use one of these two maps to correct for RM variations, the variance of the white noise can be conservatively set to 0.06--0.07 rad m$^{-2}$ for observations taken in Europe after 1-day sin waves and linear trend have been taken into account. This is approximately an order of the magnitude higher than the uncertainties on the observed RM, obtained from BGLSP, for the pulsars considered. As we have used the data of only three pulsars and our observational sites are located only in Germany, this value can vary, e.g., increasing significantly in places with sparse GNSS station coverage. Thus, in order to get reliable estimates of the sensitivity to long-term RM variations for a specific instrument, we recommend to undertake a similar kind of analysis for their sites independently.

Essentially, the determined values along with BGLSP uncertainties define the sensitivity of RM measurements to astrophysical signals. One of the promising signal of interest, when dealing with Faraday rotation studies, is the time-variable interstellar contribution to the RM. Let us assume that the ionised ISM is homogeneous. Then, the relative motion between a pulsar moving with velocity $\bm{v}$ and an observer can cause temporal RM variations induced by the change both in the projection of the magnetic field on the LoS and in the pulsar distance $L$. By differentiating Eq. (\ref{RMlambda}) under the assumption of a small change between the initial and the final position of a pulsar, we have:
\begin{equation}
\begin{split}
\Delta \text{RM} \simeq - 0.81n_{\text{e}} B \cdot v_{\perp} T \sin \theta + 0.81n_{\text{e}} B \cdot v_{\parallel} T \cos \theta \\
\sim 3\times10^{-6} \text{rad m}^{-2} \left( \frac{L}{1 \text{ kpc}} \right)^{-1} \left( \frac{\text{RM}}{30 \text{ rad m}^{-2}} \right) \left( \frac{|\bm{v_{\perp}}|}{100 \text{ km s}^{-1}} \right) \left( \frac{T}{\text{ yr}}\right)
\end{split}
\end{equation}
where $\theta$ is the angle between the magnetic field vector and the LoS, and $T$ is the whole timespan.

Besides this deterministic signal, we expect a time-variable stochastic part of the interstellar contribution, as predicted by the Kolmogorov turbulence \citep{1941DoSSR..30..301K}. As it was shown in \cite{2013MNRAS.429.2161K} the power spectral density of the stochastic contribution is $\text{PSD}_{\text{KL}} =0.0112\times D(\tau)\tau^{-\frac{5}{3}}f^{-\frac{8}{3}}$, where $D(\tau)$ is the structure function. The estimated rms of RM will increase with time $T$ \citep{1996ApJ...458..194M, 2016ApJ...824..113X} as:
\begin{equation}
\begin{split}
\text{rms}_{\text{RM}} \sim 0.81\sqrt{n_{\text{e}}^2 \sigma_{B}^2 + B_{\parallel}^2\sigma_{n}^2}L
\sim \sqrt{\int_{1/T}^{\infty}
\text{PSD}_{\text{KL}}(f) \text{d}f}\\
= 6 \times 10^{-5} \text{rad m}^{-2} \left(\frac{L}{1\text{kpc}}\right)^{\frac{1}{2}}\left(\frac{|v|}{100 \text{km s}^{-1}}\right)^{\frac{5}{6}}\left(\frac{T}{\text{yr}}\right)^{\frac{5}{6}}
\label{RM-ism_pred}
\end{split}
\end{equation}
where $\sigma_{B}^2$ and $\sigma_{n}^2$ are the variances of magnetic field and electron density fluctuation, respectively.

These calculations show that the signals of interest are characterised by a very small amplitudes, of the order of $10^{-5}$--$10^{-4}$ rad m$^{-2}$ , which is several orders of magnitude lower than the observed RM variations in this work. From the comparison of the power spectral densities\footnote{$\text{PSD}_{\text{KL}}\simeq \text{PSD}_{\text{WN}}=\sigma^2 /f_{\text{Ny}}$, where $\text{PSD}_{\text{WN}}$ is the power spectral density of the white noise and $f_{\text{Ny}}$ is the Nyquist frequency of our dataset.} we can conclude that we need $\sim$40 years of observations with the current sensitivity (mostly limited by the imperfections of the ionospheric modelling) for this kind of signals to become significant.

More promising signals of astrophysical nature could be registered thanks to extreme scattering events \citep{2015ApJ...808..113C}, associated with the passage of a blob of high density plasma through the LoS, extreme magneto-ionic environment of the source \citep{2018ApJ...852L..12D}, and coronal mass ejections \cite{2016ApJ...831..208H}, which may cause more prominent RM perturbations. 

A deeper understanding of the physics of ionospheric behaviour and instrumental GPS biases, along with the development of more regular GPS station arrays in the direct vicinity to the radio telescopes will improve the quality of the estimates of TEC in the ionospheric layer, which will, in turn, increase our sensitivity to the astrophysical RM variations.

\section*{Acknowledgements}
The authors would like to thank William Coles and Nicolas Caballero for fruitful discussions. NP acknowledges the support from IMPRS Bonn/Cologne and the Bonn-Cologne Graduate School (BCGS). S.O. acknowledges the Australian Research Council grant Laureate Fellowship FL150100148.

This paper is based on data obtained with the German
stations constructed by ASTRON International LOFAR Telescope (ILT) \citep{2013A&A...556A...2V} under project
codes LC0\textunderscore014, LC1\textunderscore048, LC2\textunderscore011, LC3\textunderscore029, LC4\textunderscore025, LT5\textunderscore001 (run until cycle 8), LC9\textunderscore039 and during
station-owners time. LOFAR \citep{2013A&A...556A...2V} is the Low Frequency Array designed and constructed by ASTRON. In this work we made use of data from the Effelsberg (DE601) LOFAR station funded by the Max-Planck--Gesellschaft; the Tautenburg (DE603) LOFAR station funded by the State of Thuringia, supported by the European Union (EFRE) and the Federal Ministry of Education and Research (BMBF), Verbundforschung project D-LOFAR I (grant 05A08ST1); the J\"ulich (DE605) LOFAR station supported by the BMBF Verbundforschung project D-LOFAR I (grant 05A08LJ1); and the Norderstedt (DE609) LOFAR station funded by the BMBF Verbundforschung project D-LOFAR II (grant 05A11LJ1). The observations of the German LOFAR stations were carried out in the stand-alone GLOW mode (German LOng-Wavelength array), which is technically operated and supported by the Max-Planck-Institut f\"ur Radioastronomie, the Forschungszentrum J\"ulich and Bielefeld University. We acknowledge support and operation of the GLOW network, computing and storage facilities by the FZ J\"ulich, the MPIfR and Bielefeld University and financial support from 
BMBF D-LOFAR III (grant 05A14PBA), and by the states of Nordrhein-Westfalia and Hamburg.

This research has made extensive use of NASA's Astrophysics Data System and the ATNF Pulsar Catalog.




\bibliographystyle{mnras}
\bibliography{iono_pulsar.bib} 




\appendix

\section{Classical RM synthesis technique and the ways to improve it}
\label{app:rm} 

Here we inspect the Faraday rotation induced by a magnetoionic medium on the radiation of a point source, in the case 
we are not affected by the effects of multibeam propagation such as differential Faraday rotation and wavelength-dependent polarization \citep{1998MNRAS.299..189S}.
As mentioned in Section~\ref{sec:intro}, the induced variation in the polarization angle is proportional to the square of the observational wavelength $\lambda$.
It can be shown (see Eqs. below) that a nonzero RM gives rise to harmonic signals in both of the Stokes parameters $Q$ and $U$ across the bandwidth, which are shifted by $\pi/2$ with respect to each other. This allows us to estimate not only the absolute value of RM, but also its sign through the RM synthesis and related techniques \citep{1966MNRAS.133...67B, 2005A&A...441.1217B}. These algorithms can recover the harmonic signal in RM ranges from $\pi/[\lambda_{\text{max}}^2-\lambda_{\text{min}}^2]$ up to approximately average Nyquist boundary $~\pi/2\delta (\lambda^2)$, where  $\delta (\lambda^2)$ is average size of the sample, determined by the size of the frequency channel, and $\lambda_{\text{min}}$ and $\lambda_{\text{max}}$ are the lowest and highest observational wavelengths, respectively. In our case, the range of available RMs is $~0.5$--$120\text{ rad m}^{-2}$. As we are working in a relatively low RM regime, the effects of depolarization in the frequency channels are neglected  in this work \citep{2015MNRAS.447L..26S}.

Mathematically, the problem of RM search can be described in the following way. The expected Stokes $Q_{\text{mod}}$ and $U_{\text{mod}}$ can be expressed as functions of the intrinsic intensity $I_{\text{mod}}$, polarization fraction $p$, and angles $\psi$ and $\chi$, which characterize the polarization ellipse,
\begin{equation}
\begin{split}
Q_{\text{mod}}=I_{\text{mod}}p\cos 2\chi \cos 2 \psi=I_{\text{mod}}p\cos 2\chi \cos(2 \text{RM}\lambda^2+2\psi_0),\\
U_{\text{mod}}=I_{\text{mod}}p\cos 2\chi \sin 2 \psi=I_{\text{mod}}p\cos 2\chi \sin(2 \text{RM}\lambda^2+2\psi_0).
\end{split}
\end{equation}
In practice, we can only access measured Stokes parameters $I_{\text{obs}}$, $Q_{\text{obs}}$, $U_{\text{obs}}$, corrupted by noise. For sake of simplicity, as we do not know the intrinsic intensity of the source because we do not perform flux calibration, we use $Q_{\text{obs}}/I_{\text{obs}}$ and $U_{\text{obs}}/I_{\text{obs}}$, denoted as $q$ and $u$, respectively. In this work, we assume that $q$ and $u$ are distributed normally around their mean values with variances:
\begin{equation}
\begin{split}
\sigma_{q}=\frac{Q}{I}\sqrt{\left(\frac{\sigma_Q}{Q}\right)^2+\left(\frac{\sigma_I}{I}\right)^2},\\
\sigma_{u}=\frac{U}{I}\sqrt{\left(\frac{\sigma_U}{U}\right)^2+\left(\frac{\sigma_I}{I}\right)^2},
\end{split}
\end{equation}
where $\sigma_I$, $\sigma_Q$ and $\sigma_U$ are the standard deviations of the observed Stokes parameters in the off-pulse region (see Appendix \ref{gauss}).

To recover the RM, we apply to the more classical RM synthesis technique the method of the Bayesian Generalised Lomb-Scargle Periodogram (BGLSP), described in \citep{2015A&A...573A.101M, 2009A&A...496..577Z}. 

By writing $c_i=\cos(2 \text{RM}\lambda_i^2-\theta)$ and $s_i=\sin(2 \text{RM}\lambda_i^2-\theta)$, the normalised Stokes $q_{\text{obs}}$ and $u_{\text{obs}}$ can be expressed as
\begin{align}
\begin{split}
&q_{\text{obs},i}=Ac_i+Bs_i+\gamma_q+\epsilon_{q, i}, \text{and} \\
&u_{\text{obs},i}=As_i-Bc_i+\gamma_u+\epsilon_{u, i},
\end{split}
\end{align}
where $A$ and $B$ are the amplitudes of oscillation, $\gamma_q$ and $\gamma_u$ are the constant offsets associated with the instrumental peak, and $\theta$ is an arbitrary phase reference point, which does not affect the $\text{RM}$ and is defined in Appendix \ref{app:BayesLomb}. The noise contributions $\epsilon_{q, i}$ and $\epsilon_{u, i}$ are assumed to be normally distributed, with standard deviations $\sigma_{q, i}$ and $\sigma_{u, i}$ and to vary independently across frequency channels. According to Bayes' theorem, the posterior probability can be written as
\begin{equation}
P_{\text{pst}}(\text{parameters}|\text{data})=\frac{P_{\text{pr}}(\text{parameters}) P(\text{data}|\text{parameters})}{P(\text{data})},
\end{equation}
where $P_{pr}(\text{parameters})$ is the prior distribution of the unknown parameters, $P(\text{data}|\text{parameters})$ is the likelihood function, $P(\text{data})$ is the so-called Bayesian evidence, which is a normalization factor in our case and plays an important role in the problem of model selection. Assuming uniform prior distributions of the parameters, the posterior probability is proportional to a likelihood :
\begin{align}
\begin{split}
\label{eq:posterfull}
P_{\text{posterior}}(A, B, \gamma_q, \gamma_u, \text{RM}, \eta \vert q_{\text{obs}, i}, u_{\text{obs}, i})\propto \\
\Pi_{i=1}^{N_{\text{ch}}}\frac{1}{\sqrt{2\pi}\sigma_{q,i} \eta}\exp\left(-\frac{(q_{\text{obs},i}-Ac_i-Bs_i-\gamma_q)^{2}}{2(\sigma_{q,i} \eta)^2}\right)\times \\
\times \Pi_{i=1}^{N_{\text{ch}}}\frac{1}{\sqrt{2\pi}\sigma_{u,i} \eta}\exp\left(-\frac{(u_{\text{obs},i}+As_i-Bc_i-\gamma_u)^{2}}{2(\sigma_{u, i} \eta)^2}\right).
\end{split}
\end{align}
The resultant form for the posterior probability is analytically marginalised over the nuisance parameters [$A$, $B$, $\gamma_q$, $\gamma_u$] (see Eq.~(\ref{Ppost})) and is provided in Appendix \ref{app:BayesLomb}.

As the integration time of pulsar observations is not infinitely small, the rate of change of ionospheric RMs during the integration time will introduce an additional ambiguity to measured RMs, which we have taken into account here by introducing the parameter $\eta$ in the denominator of Eq. (\ref{eq:posterfull}) \citep[see also][]{2017MNRAS.466..378S}. It acts effectively as a multiplier for all the $Q$ and $U$ error bars (see Section~\ref{app:BayesLomb}) and is correlated with the reduced $\chi^2$ value. Particularly, for our case of a 15-min integration time, $\eta$ typically varies between 1.5 and 3. The parameter $\eta$ was estimated separately, using Eq.~(\ref{eta}), and fixed to its maximum likelihood value.

By performing a 1D grid search in the RM parameter space, we can successfully recover the RM posterior probability (see Fig. \ref{fig:RMsynt}). The uncertainty in the RM value is determined as the variance of the normal Gaussian distribution, fit to the resultant shape of the posterior probability.

\section{Bayesian Lomb-Scargle Periodogram}
\label{app:BayesLomb}
Here we provide the derivation of the marginalised posterior probability $P_{\text{posterior}}$ from Eq.~(\ref{eq:posterfull}). Using similar notations to \citet{2015A&A...573A.101M}, we can determine the part of the expression for the posterior probability which depends on unknown parameters (called Sufficient Statistics) as:

\begin{equation}
\begin{split}
\ln P_{\text{posterior}}\left(A, B, \gamma_q, \gamma_u, \text{RM}\vert q_{\text{obs}, i}, u_{\text{obs}, i}\right)\propto\\
-\frac{1}{2}\sum_{i=1}^{N_{\text{ch}}}\left[\frac{\left(q_{\text{obs},i}-Ac_i-Bs_i-\gamma_q\right)^2}{\sigma_{q,i}^2}+
\frac{\left(u_{\text{obs},i}+As_i-Bc_i-\gamma_u\right)^2}{\sigma_{u, i}^2}\right]= \\ 
\frac{1}{2}(-\hat{YY}+2A\hat{YC}+2B\hat{YS}+2\gamma_{q}Y_q+2\gamma_{u}Y_u-A^2\hat{CC}-\\
-B^2\hat{SS}-\gamma_{q}^2W_q-\gamma_{u}^2W_u
-2A\gamma_{q}C_q-2A\gamma_{u}C_u
-2B\gamma_{q}S_{q}\\
-2B\gamma_{u}S_{u}).
\end{split}
\end{equation}

The cross term $AB\sum_{i=1}^{N_{\text{ch}}}\left(\omega_{q,i}-\omega_{u,i}\right)c_i s_i$ can be suppressed by assuming that $\tan(2\theta)=\sum_{i=1}^{N_{\text{ch}}}\left(\omega_{q,i}-\omega_{u,i}\right)\sin\left(4\text{RM}\lambda^2\right)/\sum_{	i=1}^{N_{\text{ch}}}\left(\omega_{q,i}-\omega_{u,i}\right)\cos\left(4\text{RM}\lambda^2\right)$.
In the above expression the following denominations were used:
\begin{align}
\begin{split}
W_q &=\sum_{i=1}^N \omega_{q,i}, \text{and} \ W_u=\sum_{i=1}^N \omega_{u,i},\\
Y_q &=\sum_{i=1}^N \omega_{q,i} q_{\text{obs},i}, \text{and} \ Y_u=\sum_{i=1}^N \omega_{u,i} u_{\text{obs},i},\\
\hat{YY} &=\sum_{i=1}^N \omega_{q,i} q_{\text{obs}, i}^2+ \omega_{u,i} u_{obs, i}^2,\\
\hat{YC} &=\sum_{i=1}^N \omega_{q,i} q_{\text{obs}, i} c_i+ \omega_{u,i} u_{obs, i} s_i,\\
\hat{YS} &=\sum_{i=1}^N \omega_{q,i} q_{\text{obs}, i} s_i- \omega_{u,i} u_{obs,i} c_i,\\
\hat{CC} &=\sum_{i=1}^N \omega_{q,i} c^2_i+ \omega_{u,i} s^2_i, \hat{SS}=\sum_{i=1}^N \omega_{q,i} s^2_i+ \omega_{u,i} c^2_i,\\
C_q &=\sum_{i=1}^N \omega_{q,i} c_i, \text{and} \ C_u=\sum_{i=1}^N \omega_{u,i} c_i,\\
S_q &=\sum_{i=1}^N \omega_{q,i} s_i  \text{ and} \ S_u=\sum_{i=1}^N \omega_{q,i} s_i,
\end{split}
\end{align}
and the weights are defined in a traditional way as:
\begin{equation}
\begin{split}
\omega_{q,i}=\frac{1}{\sigma^{2}_{q,i}} \text{ and } \omega_{u,i}=\frac{1}{\sigma^{2}_{u,i}}.
\end{split}
\end{equation}
The resultant expression for the Sufficient Statistics after marginalization over nuisance parameters {$A, B, \gamma_{q}, \gamma_{u}$} is
\begin{equation}
\begin{split}
P_{\text{posterior}}\left(\text{RM} \vert q_{\text{obs}}, u_{\text{obs}}\right) \propto \\
\frac{1}{\sqrt{|4DF-E^2|\hat{CC}\hat{SS}}}\exp\left(M-\frac{\hat{YY}}{2}+\frac{DG^2-EGJ+FJ^2}{E^2-4DF}\right),
\label{Ppost}
\end{split}
\end{equation}
where
\begin{align}
\begin{split}
D &=\frac{C_q^2 \hat{SS}+ S_q^2 \hat{CC} -W_q \hat{CC} \hat{SS}}{2\hat{CC}\hat{SS}},\\
F &=\frac{C_u^2 \hat{SS}+ S_u^2 \hat{CC} -W_u \hat{CC} \hat{SS}}{2\hat{CC}\hat{SS}},\\
E &=\frac{C_u C_q \hat{SS}+S_u S_q \hat{CC}}{\hat{CC}\hat{SS}},\\
J &=\frac{C_q \hat{YC} \hat{SS} + S_q \hat{YS} \hat{CC}-Y_q \hat{CC} \hat{SS}}{\hat{CC}\hat{SS}},\\
G &=\frac{C_u \hat{YC} \hat{SS} + S_u \hat{YS} \hat{CC}-Y_u \hat{CC} \hat{SS}}{\hat{CC}\hat{SS}} \text{ and } \\
M &=\frac{\hat{YC}^2\hat{SS}+\hat{YS}^2\hat{CC}}{2\hat{CC}\hat{SS}}.
\end{split}
\end{align}
The resultant expression can be easily generalised to the case of underestimated uncertainties in Stokes $Q$ and $U$ by including an extra free parameter $\eta$, such that $\omega_{q} \rightarrow \eta^{-2} \omega_{q}$ and $\omega_{u} \rightarrow \eta^{-2} \omega_{u}$. In this case the resultant marginalised posterior probability will be the function of two parameters, $P_{\text{posterior}}\left(\text{RM}, \eta|q_{\text{obs}},u_{\text{obs}}\right)$. In order to account properly for the fuzzy structure around 0 rad m$^{-2}$, a more complex model of the noise should be used, e.g., a power-law red noise, which is not under consideration in this paper. 

In order to determine the unknown parameters within the Bayesian framework, one needs to numerically reconstruct the 2D posterior probability, which can be effectively managed by the Markov Chain Monte Carlo. In the frequentist approach, which is less computationally expensive and was used in this paper, we are interested in the maximum likelihood estimation of the unknown parameters, which for $\eta$ can be found analytically:
\begin{equation}
\hat{\eta}^2=-\frac{2}{2N_{\text{ch}}-4}\left[M-\frac{\hat{YY}}{2}+\frac{DG^2-EGJ+FJ^2}{E^2-4DF}\right].
\label{eta}
\end{equation}

\section{Notes on the distribution of Q/I and U/I}
\label{gauss}
A full investigation of the Gaussianity of the $q$ and $u$ distributions is beyond the scope of this paper. However, here we will have some general comment on this.

\begin{figure}
\includegraphics[width=1.1\columnwidth]{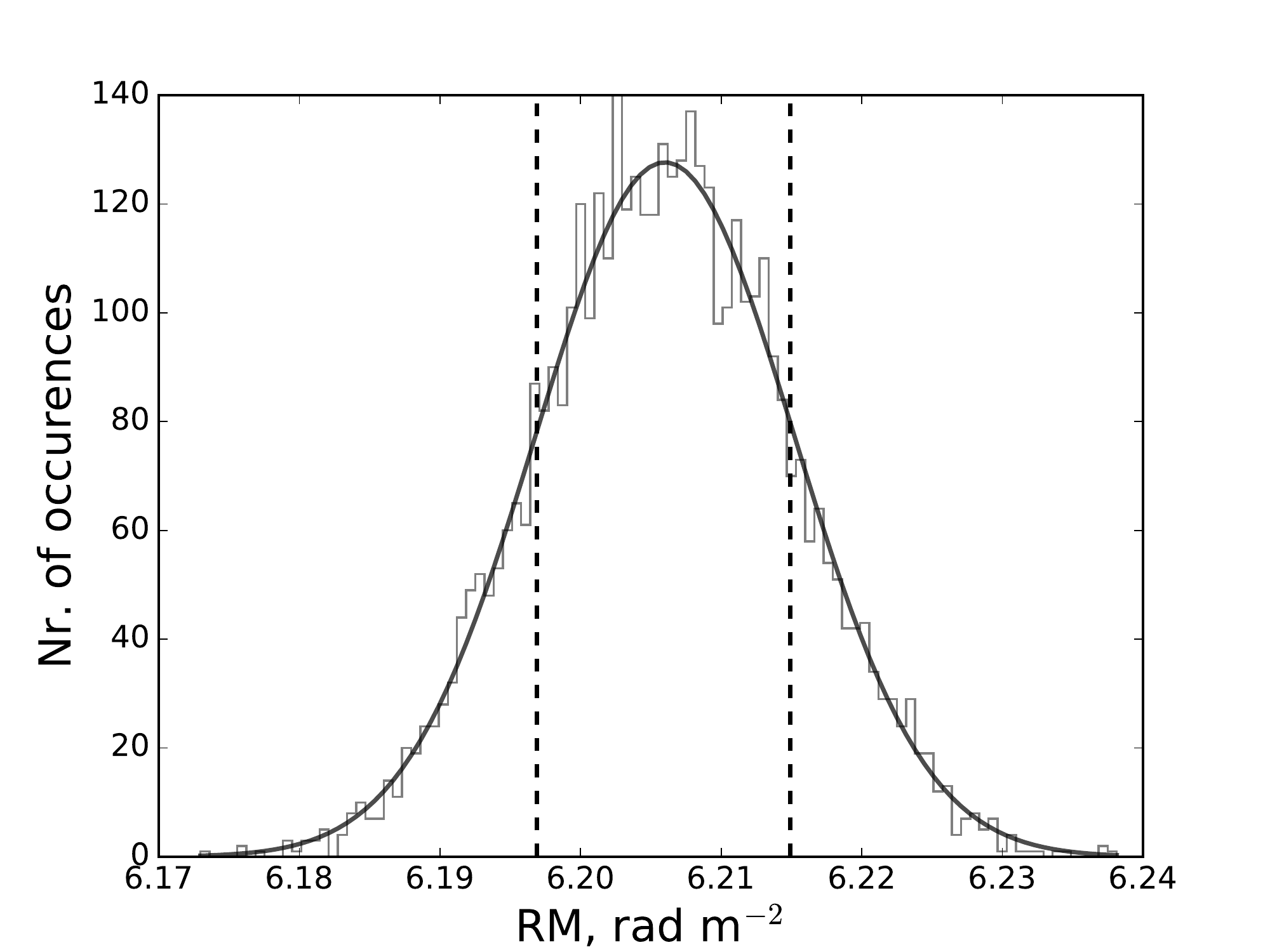}
\label{pic:gist}
\caption{Distribution of the reconstructed RMs for simulated data of PSR J1136+1551 (grey line) and posterior probability of RM as predicted with BGLSP method (black line). The vertical dashed lines show the quantiles of the reconstructed distribution $Q_{16\%}$ and $Q_{84\%}$. The half of the interquartile range of the RM distribution, reconstructed from the simulations, is $(Q_{84\%}-Q_{16\%})/2=0.0090$ rad m$^{-2}$, while the $1\sigma$ uncertainty, determined with the BGLSP is $0.0092$ rad m$^{-2}$. The S/N of PSR J1136+1551 in linear intensity is $\simeq$50.}
    \label{fig:gist}
\end{figure}

By postulating that the observed Stokes $I_{\text{obs}}$, $Q_{\text{obs}}$ and $U_{\text{obs}}$ are normally distributed, one can derive that $q$ and $u$ will actually follow a Cauchy-like distributions. In \citet{2017MNRAS.466..378S} it was showed that the non-Gaussianity of $q$ and $u$ can potentially bias the uncertainty of measured RM in the low S/N regime\footnote{ \citet{2017MNRAS.466..378S} demonstrated that this problem can be avoided by introducing $N_{\text{ch}}$ (number of frequency channels) nuisance parameters $I_{\text{mod}, i}$, and found the analytical expression for the likelihood, marginalised over these parameters in the case of weakly polarised sources ($L_{\text{mod}}<<I_{\text{mod}}$).}. However, in the case of high S/N (i.e., $\sigma_I/I<0.1$, see \cite{1975mnsc.21.11.1338, doi:10.1152/jappl.2000.88.6.2279}), the resultant Cauchy distribution can be reasonably well approximated by the normal distribution. By selecting frequency channels above the threshold, and simulating the normally distributed Stokes parameters in each of them, we have reconstructed the RM distribution determined with the BGLSP method. For the two pulsars that we have included in the test (PSRs J0332+5434 and J1136+1551), we have found that the resultant RM distribution can be well approximated by a normal one and its parameters (variance and mean) are in good agreement with those determined with the BGLSP. In Fig. \ref{pic:gist} we display the RM distribution for PSR J1136+1551. The properties if the reconstructed distribution are given in caption.

The further analysis of non-Gaussianity of $q$ and $u$ and its influence on the distribution of the resultant RMs will be addressed in future work.


\section{Results of the analysis of the systematics for the other geomagnetic models}

\begin{table*}
	\centering
	\caption{ Estimation of the noise parameters based on the Bayesian analysis of RM residuals using EMM geomagnetic model and different ionospheric maps. The results are presented in the same fashion as in Table 2 of the paper. The analysis was based on datasets of PSR~J0332+5434, PSR~J1136+1551 and of PSR~J0814+7429. We report the median and 1-sigma uncertainty values of the amplitude of the Lorentzian spectrum $A_\text{L}$ [rad m$^{-2}$], which effectively quantifies the measurement uncertainties of the ionospheric RM corrections; the turnover frequency of the Lorentzian spectrum $f_0$ [day$^{-1}$], the amplitude of the 1-day harmonic signal $A_\text{d}$ [rad m$^{-2}$] and maximum likelihood estimation of $\text{RM}_{\text{IISM}}$ [rad m$^{-2}$]. The latter is assumed to be constant across a timescale of several months. The factor $f_{B}=1.11$ was applied.}
	\label{tab:example_table2}
	\begin{tabular}{lccccccr} 
		\hline
        &&PSR J0332+5434\\
        \hline
		Model & $A_\text{L}^{\text{med}}$ & $f_0^{\text{med}}$ & $A_\text{d}^{\text{med}}$ & $\text{RM}_{\text{IISM}}^{\text{ML}}$ \\
		\hline
        \hline
		UQRG &  $0.045_{-0.003}^{+0.003}$ &  $1.5_{-0.1}^{+0.2}$ & $0.012_{-0.006}^{+0.007}$ & -64.16 \\
        \\
		JPLG &  $0.051_{-0.002}^{+0.003}$ &  $1.8_{-0.2}^{+0.2}$ & $0.025_{-0.005}^{+0.007}$ & -64.21\\
	        \\
		EHRG & $0.054_{-0.003}^{+0.003}$ &  $1.2_{-0.1}^{+0.1}$ &  $0.012_{-0.008}^{+0.007}$ & -64.06\\
         \\
          IGSG & $0.060_{-0.003}^{+0.003}$ &  $1.2_{-0.1}^{+0.1}$ & $0.02_{-0.006}^{+0.005}$ & -64.08\\
        \\
         ESAG & $0.068_{-0.005}^{+0.005}$  & $1.2_{-0.1}^{+0.1}$ & $0.025_{-0.009}^{+0.007}$ & -64.07\\
         \\
        UPCG & $0.073_{-0.004}^{+0.003}$ &  $0.9_{-0.1}^{+0.1}$ & $0.025_{-0.014}^{+0.008}$ & -64.17 \\
        \\
        CODG & $0.11_{-0.01}^{+0.01}$ & $0.29_{-0.06}^{+0.06}$ & $0.063_{-0.009}^{+0.009}$ & -63.95\\
        \hline
	\end{tabular}
    \\
     \begin{tabular}{lccccccr} 
		\hline
        &&PSR J1136+1551\\
        \hline
		Model & $A_\text{L}^{\text{med}}$ & $f_0^{\text{med}}$ & $A_\text{d}^{\text{med}}$ & $\text{RM}_{\text{IISM}}^{\text{ML}}$ \\
		\hline
        \hline
		UQRG & $0.061_{-0.004}^{+0.005}$ & $2.1_{-0.3}^{+0.3}$ & $0.079_{-0.009}^{+0.01}$ &  4.16 \\
        \\
		JPLG & $0.073_{-0.004}^{+0.004}$ & $1.9_{-0.5}^{+0.6}$ & $-$ & 4.02 \\
        \\
        EHRG & $0.082_{-0.004}^{+0.005}$ & $1.3_{-0.2}^{+0.2}$ & $0.03_{-0.01}^{+0.01}$ & 4.22 \\
        \\
        IGSG & $0.137_{-0.008}^{+0.009}$ & $0.6_{-0.1}^{+0.1}$ & $0.04_{-0.02}^{+0.02}$ & 4.17 \\
        \\
        ESAG & $0.116_{-0.007}^{+0.008}$ & $1.0_{-0.1}^{+0.1}$ & $0.03_{-0.01}^{+0.02}$ & 4.27 \\
        \\
        UPCG & $0.127_{-0.008}^{+0.01}$ & $0.9_{-0.1}^{+0.1}$  & $0.08_{-0.02}^{+0.02}$ & 4.18 \\
        \\
        CODG & $0.22_{-0.01}^{+0.02}$ & $0.14_{-0.03}^{+0.03}$ &   $0.08_{-0.02}^{+0.02}$  & 4.18\\
		\hline
	\end{tabular}
    \\
	\begin{tabular}{lccccccr} 
		\hline
        &&PSR J0814+7429\\
        \hline
		Model  & $A_\text{L}^{\text{med}}$ & $f_0^{\text{med}}$ &  $A_\text{d}^{\text{med}}$ & $\text{RM}_{\text{IISM}}^{\text{ML}}$ \\
		\hline
        \hline
        UQRG & $0.053_{-0.003}^{+0.004}$ & $2.8_{-0.5}^{+0.5}$ & $0.052_{-0.007}^{+0.006}$ &  -13.75 \\
        \\
        JPLG & $0.051_{-0.002}^{+0.004}$ &$2.4_{-0.4}^{+0.4}$ &   $0.024_{-0.008}^{+0.006}$ & -13.79 \\
        \\
        EHRG & $0.054_{-0.003}^{+0.003}$ & $2.1_{-0.3}^{+0.3}$ & $0.033_{-0.02}^{+0.01}$ & -13.66 \\
        \\
        IGSG & $0.064_{-0.004}^{+0.005}$ & $1.5_{-0.3}^{+0.3}$ &  $0.051_{-0.008}^{+0.01}$ & -13.69 \\
        \\
        ESAG & $0.068_{-0.004}^{+0.005}$  & $1.5_{-0.3}^{+0.3}$ 	&  $0.03_{-0.01}^{+0.01}$ & -13.65 \\
        \\
         UPCG & $0.069_{-0.006}^{+0.005}$  & $1.4_{-0.3}^{+0.3}$ &  $0.045_{-0.01}^{+0.01}$ & -13.74 \\
        \\
        CODG & $0.09_{-0.01}^{+0.01}$ & $0.8_{-0.1}^{+0.1}$ &  $0.07_{-0.01}^{+0.02}$  & -13.64\\
		\hline
	\end{tabular}
\end{table*}

\begin{table*}
	\centering
	\caption{Analogous to Table \ref{tab:example_table1} and \ref{tab:example_table2}, but with IGRF geomagnetic model used for ionospheric RM modeling.}
	\label{tab:example_table3}
	\begin{tabular}{lccccccr} 
		\hline
        &&PSR J0332+5434\\
        \hline
		Model & $A_\text{L}^{\text{med}}$ & $f_0^{\text{med}}$ & $A_\text{d}^{\text{med}}$ & $\text{RM}_{\text{IISM}}^{\text{ML}}$ \\
		\hline
        \hline
		UQRG &  $0.045_{-0.002}^{+0.002}$ &  $1.5_{-0.1}^{+0.2}$ & $0.012_{-0.006}^{+0.007}$ & -64.16 \\
        \\
		JPLG &  $0.049_{-0.002}^{+0.002}$ &  $1.7_{-0.3}^{+0.3}$ & $0.025_{-0.005}^{+0.007}$ & -64.21\\
	        \\
		EHRG & $0.054_{-0.003}^{+0.003}$ &  $1.2_{-0.1}^{+0.1}$ &  $0.012_{-0.008}^{+0.007}$ & -64.05\\
         \\
          IGSG & $0.060_{-0.003}^{+0.003}$ &  $1.2_{-0.1}^{+0.1}$ & $0.02_{-0.007}^{+0.005}$ & -64.09\\
        \\
         ESAG & $0.068_{-0.004}^{+0.005}$  & $1.1_{-0.1}^{+0.1}$ & $0.025_{-0.009}^{+0.007}$ & -64.05\\
         \\
        UPCG & $0.073_{-0.004}^{+0.003}$ &  $0.9_{-0.1}^{+0.1}$ & $0.025_{-0.014}^{+0.008}$ & -64.17 \\
        \\
        CODG & $0.11_{-0.01}^{+0.01}$ & $0.29_{-0.06}^{+0.06}$ & $0.063_{-0.009}^{+0.009}$ & -63.95\\
        \hline
	\end{tabular}
    \\
     \begin{tabular}{lccccccr} 
		\hline
        &&PSR J1136+1551\\
        \hline
		Model & $A_\text{L}^{\text{med}}$ & $f_0^{\text{med}}$ & $A_\text{d}^{\text{med}}$ & $\text{RM}_{\text{IISM}}^{\text{ML}}$ \\
		\hline
        \hline
		UQRG & $0.061_{-0.004}^{+0.005}$ & $2.3_{-0.3}^{+0.3}$ & $0.079_{-0.009}^{+0.01}$ &  4.16 \\
        \\
		JPLG & $0.073_{-0.004}^{+0.004}$ & $1.8_{-0.5}^{+0.6}$ & $-$ & 4.03 \\
        \\
        EHRG & $0.082_{-0.004}^{+0.005}$ & $1.3_{-0.2}^{+0.2}$ & $0.03_{-0.01}^{+0.02}$ & 4.22 \\
        \\
        IGSG & $0.141_{-0.009}^{+0.009}$ & $0.6_{-0.1}^{+0.1}$ & $0.04_{-0.02}^{+0.02}$ & 4.17 \\
        \\
        ESAG & $0.114_{-0.007}^{+0.008}$ & $1.0_{-0.1}^{+0.1}$ & $0.04_{-0.01}^{+0.02}$ & 4.26 \\
        \\
        UPCG & $0.125_{-0.008}^{+0.010}$ & $0.9_{-0.1}^{+0.2}$  & $0.08_{-0.02}^{+0.02}$ & 4.17 \\
        \\
        CODG & $0.21_{-0.02}^{+0.02}$ & $0.12_{-0.03}^{+0.03}$ &   $0.06_{-0.02}^{+0.02}$  & 4.18\\
		\hline
	\end{tabular}
\\
	\begin{tabular}{lccccccr} 
		\hline
        &&PSR J0814+7429\\
        \hline
		Model  & $A_\text{L}^{\text{med}}$ & $f_0^{\text{med}}$ &  $A_\text{d}^{\text{med}}$ & $\text{RM}_{\text{IISM}}^{\text{ML}}$ \\
		\hline
        \hline
        UQRG & $0.052_{-0.003}^{+0.004}$ & $2.4_{-0.5}^{+0.5}$ & $0.048_{-0.007}^{+0.006}$ &  -13.75 \\
        \\
        JPLG & $0.051_{-0.003}^{+0.004}$ &$2.3_{-0.4}^{+0.4}$ &   $0.024_{-0.008}^{+0.006}$ & -13.79 \\
        \\
        EHRG & $0.054_{-0.003}^{+0.004}$ & $2.1_{-0.3}^{+0.3}$ & $0.035_{-0.02}^{+0.01}$ & -13.67 \\
        \\
        IGSG & $0.064_{-0.004}^{+0.005}$ & $1.4_{-0.3}^{+0.3}$ &  $0.05_{-0.008}^{+0.01}$ & -13.69 \\
        \\
        ESAG & $0.067_{-0.005}^{+0.005}$  & $1.6_{-0.3}^{+0.3}$ 	&  $0.03_{-0.01}^{+0.01}$ & -13.65 \\
        \\
         UPCG & $0.068_{-0.005}^{+0.005}$  & $1.4_{-0.3}^{+0.3}$ &  $0.045_{-0.01}^{+0.01}$ & -13.74 \\
        \\
        CODG & $0.10_{-0.01}^{+0.01}$ & $0.7_{-0.1}^{+0.1}$ &  $0.07_{-0.01}^{+0.02}$  & -13.63\\
		\hline
	\end{tabular}
\end{table*}

\bsp	
\label{lastpage}
\end{document}